\documentclass[fleqn,10pt]{wlscirep}
\usepackage[utf8]{inputenc}
\usepackage[T1]{fontenc}
\usepackage{graphicx}
\usepackage{pythonhighlight}
\usepackage[]{amsmath}
\usepackage{amssymb}
\usepackage{enumerate}
\usepackage{array}
\usepackage{multirow}
\usepackage{wrapfig}
\usepackage{fullpage}
\usepackage{overpic}
\usepackage{verbatim}
\usepackage{soul}
\title{Machine learning of pair-contact process with diffusion}

\author[1]{Jianmin Shen}
\author[1,*]{Wei Li}
\author[2]{Shengfeng Deng}
\author[1]{Dian Xu}
\author[1]{Shiyang Chen}
\author[1,3]{Feiyi Liu}
\affil[1]{Key Laboratory of Quark and Lepton Physics (MOE) and Institute of Particle Physics, Central China Normal University, Wuhan 430079, China}
\affil[2]{Institute of Technical Physics and Materials Science, Center for Energy Research, Budapest 1121, Hungary}
\affil[3]{Institute for Physics,E{\"o}tv{\"o}s Lor\'and University\\1/A P\'azm\'any P. S\'et\'any, H-1117, Budapest, Hungary}

\affil[*]{liw@mail.ccnu.edu.cn}

\begin{abstract}
The pair-contact process with diffusion (PCPD), a generalized model of the ordinary pair-contact process (PCP) without diffusion, exhibits a continuous absorbing phase transition. Unlike the PCP, whose nature of phase transition is clearly classified into the directed percolation (DP) universality class, the model of PCPD has been controversially discussed since its infancy. To our best knowledge, there is so far no consensus on whether the phase transition of the PCPD falls into the unknown university classes or else conveys a new kind of non-equilibrium phase transition. In this paper, both unsupervised and supervised learning are employed to study the PCPD with scrutiny. Firstly, two unsupervised learning methods, principal component analysis (PCA) and autoencoder, are taken. Our results show that both methods can cluster the original configurations of the model and provide reasonable estimates of thresholds. Therefore, no matter whether the non-equilibrium lattice model is a random process of unitary (for instance the DP) or binary (for instance the PCP), or whether it contains the diffusion motion of particles, unsupervised learning can capture the essential, hidden information. Beyond that, supervised learning is also applied to learning the PCPD at different diffusion rates. We proposed a more accurate numerical method to determine the spatial correlation exponent $\nu_{\perp}$, which, to a large degree, avoids the uncertainty of data collapses through naked eyes.
\end{abstract}
\begin{document}

\flushbottom
\maketitle
%
%
\thispagestyle{empty}


\section*{Introduction}

Machine learning (ML) algorithms \cite{jordan2015machine} have been widely used in equilibrium phase transitions, to distinguish matter phases and detect phase transitions \cite{carrasquilla2017machine,zhang2019machine,wang2016discovering,wetzel2017unsupervised,hu2017discovering,wang2017machine} in various kinds of systems. Based on whether labels are involved or not, ML methods can be categorized into supervised and unsupervised learning. Often, they are also closely related to the so-called deep learning in which more elaborate frameworks are adopted \cite{goodfellow2016deep,glassner2018deep}. As is known, supervised learning includes regression and classification, which are efficient in predicting certain quantities that appear in fields such as biophysics\cite{mckinney2006machine,schafer2014learning}, astrophysics \cite{vanderplas2012introduction}, quantum physics \cite{dunjko2018machine}, and many more domains in physics \cite{mehta2019high,carleo2019machine}. In fields such as statistical physics \cite{engel2001statistical,mehta2014exact} and condensed matter physics \cite{carrasquilla2017machine,wang2016discovering,van2017learning,carleo2017solving}, supervised learning is employed to identify phases or predict phase transitions, as well as speed up simulations \cite{shen2018self}. For supervised learning of phase transitions, we need to have real experimental data ready or generate configuration data through Monte Carlo simulations \cite{hammersley2013monte}, before labeling and training them. By this means, the trained model can recognize and predict newly input configurations and obtain the corresponding regression or classification results, from which we can also utilize the rescaling method to yield some critical exponents.

In contrast, unsupervised learning does not require labels, which in autoencoder is what we say the input itself. Unsupervised learning is powerful for data clustering, compression, dimensionality reduction and visualization, due to its ability to extract essential information from raw data. It is believed that the unsupervised learning can learn the hidden information in the input data with a changing trend, which has been intriguing.

In recent years there has been vast progress in supervised and unsupervised learning for equilibrium phase transitions \cite{carrasquilla2017machine,zhang2019machine,wang2016discovering,
wetzel2017unsupervised,hu2017discovering,wang2017machine,li2018applications} as well as non-equilibrium phase transitions \cite{bukov2018reinforcement,venderley2018machine,shen2021supervised,jo2021absorbing}. As the preprocessing for data training and prediction, unsupervised learning is more appealing. The article \cite{wang2016discovering} is the earliest literature of unsupervised learning method being used for studying phase transitions that we can retrieve so far. The author uses principal component analysis (PCA) to distinguish the phases and reveal the properties of the classical Ising model, such as order parameters and structure factors. In ref. \cite{wetzel2017unsupervised}, by using PCA, kernel-PCA, autoencoder, and variational autoencoder to study the 2-D Ising model and 3-D XY model, it is found that some potential variables could be  related to the order parameters. According to ref. \cite{hu2017discovering}, PCA has been widely applied to comparing critical behaviors of various models, including Ising models with square- and triangular-lattice, the Blume-Capel model, a highly degenerate biquadratic-exchange spin-one Ising (BSI) model, and the 2-D XY model. It is shown that the quantized principal component of PCA can not only detect phases and symmetry breaking but distinguish phase transition types and determine critical points. Similar studies of critical phenomena by using unsupervised learning can refer to \cite{zhang2019machine,scheurer2020unsupervised,wang2021unsupervised}. In what way can unsupervised learning process the known or generated data to extract essential embedded information, is also very crucial for statistical physics.

Inspired by this, the present study was designed to extrapolate ML techniques to a binary stochastic reaction process \cite{hinrichsen2000non,lubeck2004universal,henkel2008non} called pair-contact process with diffusion (PCPD) \cite{henkel2008non,odor2003critical,henkel2004non,deng2020coupled}. Two unsupervised learning methods, PCA and autoencoder, are used to study this model, respectively. PCA \cite{pearson1901liii,abdi2010principal} is a linear dimensionality reduction method, which has a wide range of applications in extracting salient features of complex data. While the autoencoder \cite{bourlard1988auto,hinton1994autoencoders,hinton2006reducing} neural network has the advantage of dealing with nonlinear data in image recognition and data compression. By applying PCA and autoencoder, we try to classify different lattice phases and extract the thresholds of the PCPD model. The input data is generated by the Monte Carlo simulations of the model.

It should be noted that the directed percolation (DP) process, a classic non-equilibrium phase transition model \cite{hinrichsen2000non,lubeck2004universal,broadbent1957percolation}, has been studied by both supervised and unsupervised learning \cite{shen2021supervised}. Combining the learning results of the DP and the PCPD models, we can immediately appreciate the value of investigating such non-equilibrium systems. Among them, the DP is a unitary random reaction process, while the PCPD is a binary one. The results show that our PCA and autoencoder can successfully cluster PCPD's configurations and determine the critical points with high accuracy. Equilibrium phase transitions have some similarities with non-equilibrium ones, and therefore similar methods are suitable for them. The unique feature of non-equilibrium phase transition is the extra dimension, namely time. The states and properties of the system may change with time. If one intends to extend the ML algorithms to non-equilibrium phase transitions, the dimension of time has to be dealt with.

The overall structure of the study takes the forms of three chapters, including this introductory one. In the second section, we briefly introduce the model of PCPD, the two unsupervised learning methods, the data sets and ML results of PCPD. Here, we successfully cluster the configurations and obtain reasonable thresholds. By using supervised learning, we also obtain some critical exponents. In addition, numerical calculations are implemented to determine correlation exponents. The third section is a summary of this paper.


\section*{Machine Learning of the PCPD}
\subsection*{The model of PCPD}
\begin{figure*}[!thb]
\begin{tabular}{ccc}
    \includegraphics[width=0.45\textwidth]{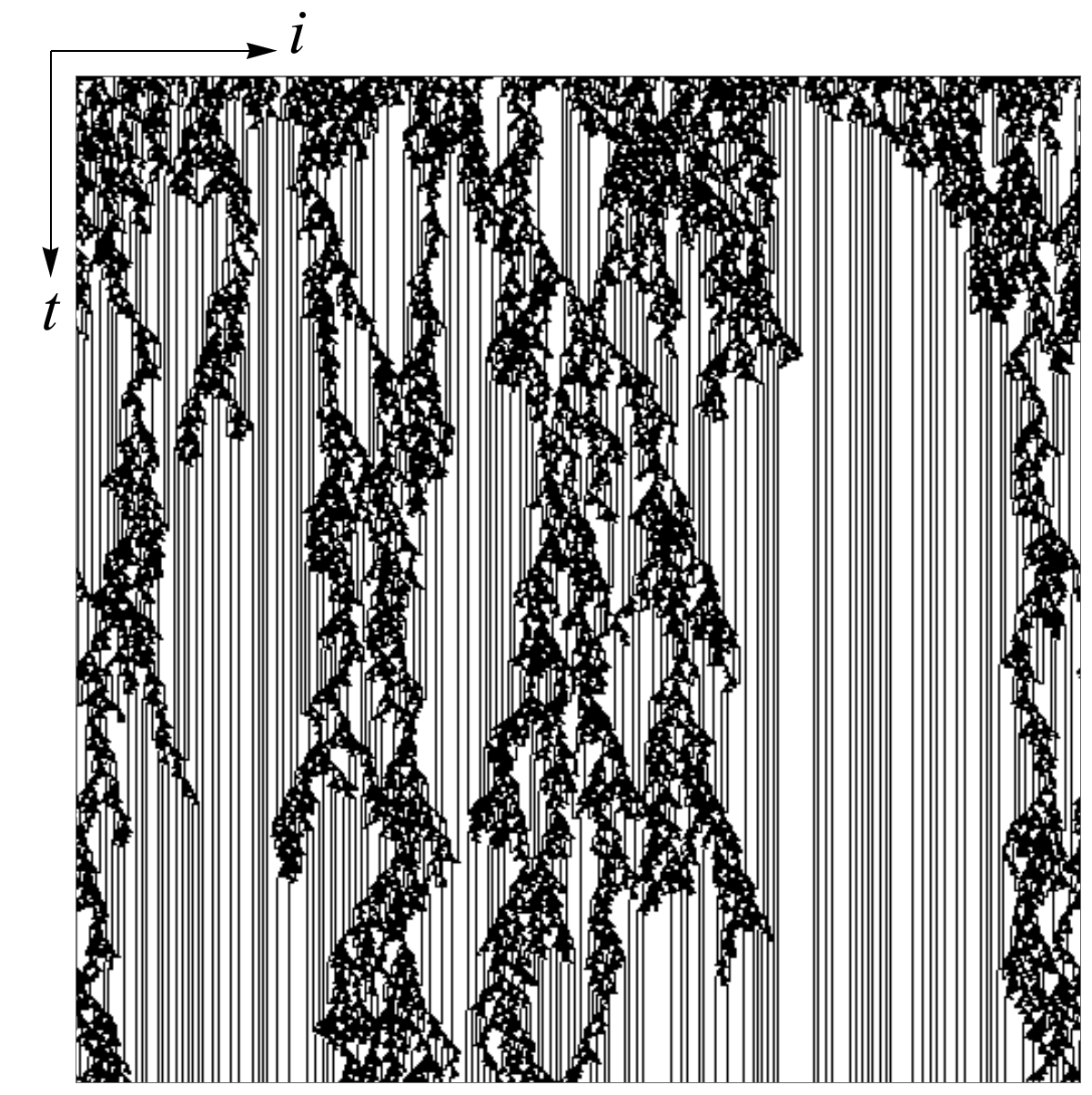} &
    \includegraphics[width=0.45\textwidth]{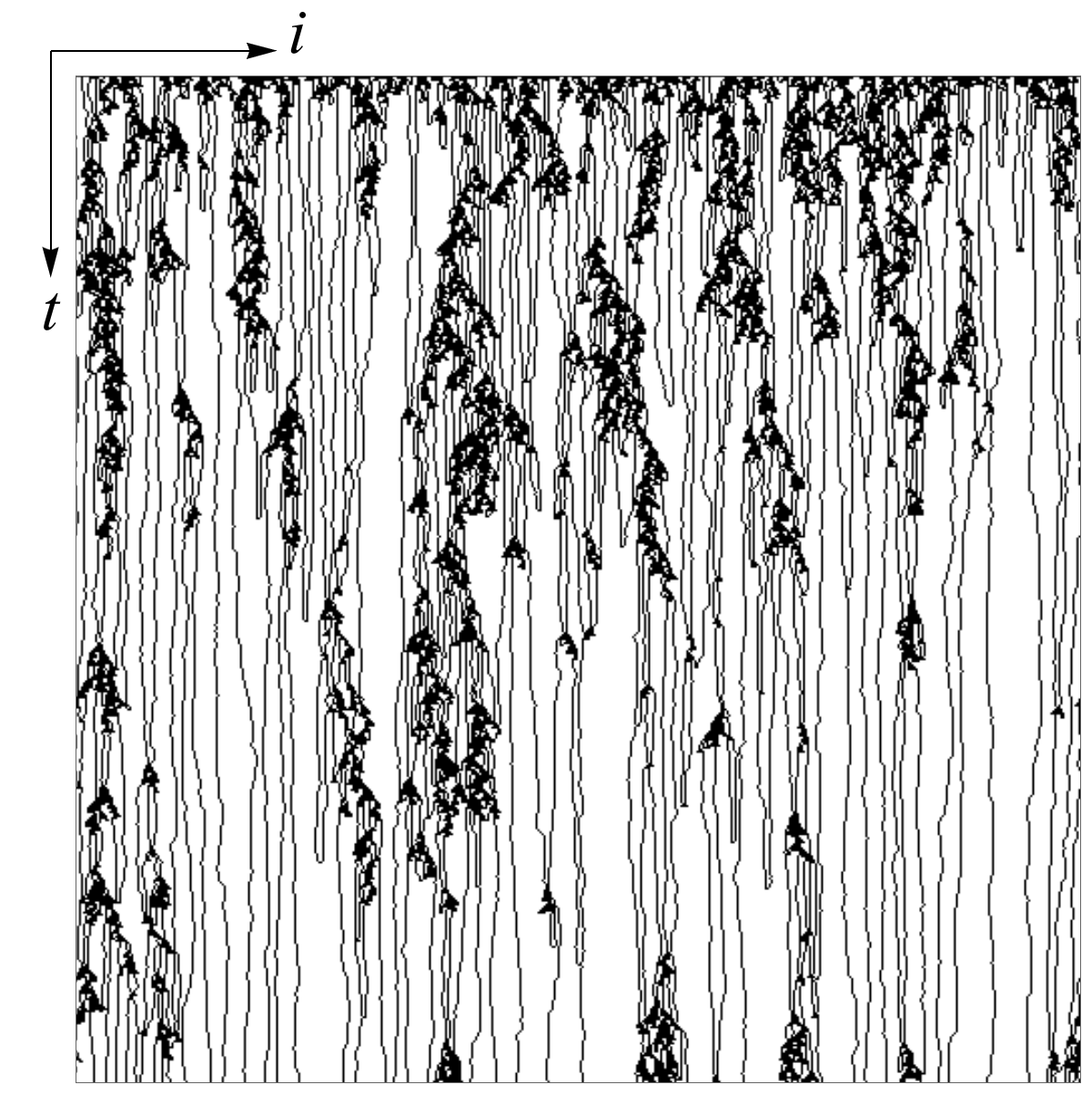} \\
     PCP &  PCPD
\end{tabular}
\caption{(1+1)-dimensional PCP and PCPD starting with a fully occupied lattice at criticality. The system size is $L = 500$, and the time step is $500$.}
\label{pcpdshow}
\end{figure*}

Before understanding the PCPD, we will briefly introduce its prototype, the pair-contact process (PCP) in which no diffusion is considered. Fig. \ref{pcpdshow} displays a configuration generated by the PCP. PCP, proposed by Jensen \cite{jensen1993critical}, is a random reaction process without particle diffusion and can produce a continuous phase transition. In the $d$-dimensional lattice, sites are either occupied or empty. Under the sequential updating mechanism, processes of proliferation and annihilation compete with each other until the system reaches a steady-state or an absorbing phase. When $L \rightarrow \infty$, the particle system has an infinite number of absorbing states. Different from the DP, the order parameter of the PCP is given by the pair-particle density. Nevertheless, the PCP has been proven to have the same critical exponents as the DP, which means that the PCP belongs to the DP universality class.

Different from the DP, the PCPD is a simple binary random reaction process \cite{grassberger1981phase}. When the diffusion rate $D$ is $0$, the classical model, PCP, is restored. Henceforth the PCPD can be regarded as a generalization of the PCP. In the PCPD, a reaction is triggered by the forming of a pair-particle. The PCPD' reaction-diffusion mechanism is given by,
\begin{equation}
 \left.\begin{aligned}
        fission &: \qquad 2A \longrightarrow 3A ,\\
       annihilation &: \qquad 2A \longrightarrow \varnothing,
       \end{aligned}
 \right.
\end{equation}
where $A$ means a single particle, and $\varnothing$ an empty site.

The PCPD demonstrates a continuous phase transition from the fluctuating active phase to the absorbing phase, whose evolution is governed by the following rules,
\begin{equation}
  \left\{
\begin{array}{rcl}
& AA \varnothing \longrightarrow AAA &; \quad with \quad rate \quad (1-p)(1-D)/2 \\
& \varnothing AA \longrightarrow AAA &; \quad with \quad rate \quad (1-p)(1-D)/2 \\
& AA \longrightarrow \varnothing \varnothing &; \quad with \quad rate \quad p(1-D) \\
& A\varnothing \longleftrightarrow \varnothing A &; \quad with \quad rate \quad D, \\
\end{array} \right.
\end{equation}
where $p \in [0,1]$ represents the pair annihilation probability, and $D \in [0,1]$ the diffusion rate. When the system is in the active phase $p < p_{c}$, the proliferation process plays a dominant role. After a sufficiently long time, the system will reach a steady-state, at which point the particle density $\rho_{s} > 0$. On the contrary, the annihilation process dominates the $p > p_{c}$ system in the absorbing phase, and the particle density decreases rapidly until the system reaches the absorbing phase. In the thermodynamics, if the diffusion rate $D$ is 0, the space-time trajectory of a single particle will always be a straight line (see left panel of Fig. \ref{pcpdshow}). Once reactions are not longer possible due to adjacent pairs being depleted, each single particle leaves behind a long stripe. In the inactive phase, since there are infinite combinations for the locations of these stripes, the process without diffusion will result in the generation of an infinite number of absorbing states. If diffusion rate $D > 0$, a single particle is allowed to diffuse, leaving a random-walk like trajectory until it collide with another particle and trigger a reaction event. Even if there are just two particles in the lattice, when the two diffusing particles meet, the offspring of them may still meet again after a long period of time, resulting in a certain degree of randomness in the trajectory of the diffusing particles, and the visual appearance are generally not straight lines. If the value of $p$ is fixed, the particle density of the system decays with the increase of the diffusion rate $D$.

In MC with periodic boundary conditions, sequential update mechanism is employed to generate configurations of the PCPD. We first select a particle and a direction randomly. Then a particle in the selected direction moves to its nearest neighboured site with probability $D$; the nearest neighbour in the selected direction annihilates with probability $p(1-D)$, and creates a new particle at the next nearest neighbourhood with probability $(1-p)(1-D)$. For a block of sites with indices $h,i,j,k,l$, we can characterize the dynamic rules of the PCPD model by Eq. \eqref{pcpd_generate_machanism}, where $s_{i}(t)$ represents the state of node $i$ at time $t$, and $z \in (0, 1) $ is a random number generated from a uniform distribution.

\begin{figure*}
\begin{equation}
\left\{
\begin{array}{rcl}
s_{j}(t+1) = 0  \quad  s_{k}(t+1) = 1   &      & {if \quad s_{j}(t) = 1 \quad and \quad s_{k}(t) = 0 \quad and \quad z < D/2,} \\
s_{j}(t+1) = 0  \quad  s_{i}(t+1) = 1  &      & {if \quad s_{j}(t) = 1 \quad and \quad s_{i}(t) = 0 \quad and \quad z < D/2,} \\
s_{j}(t+1) = 0  \quad  s_{k}(t+1) = 0  &      & {if \quad s_{j}(t) = 1 \quad and \quad s_{k}(t) = 1 \quad and \quad z < p(1-D),} \\
s_{l}(t+1) = 1  &      & {if \quad s_{j}(t) = 1 \quad and \quad s_{k}(t) = 1 \quad and \quad s_{l}(t) = 0 \quad and \quad z < (1-p)(1-D)/2,} \\
s_{h}(t+1) = 1  &      & {if \quad s_{i}(t) = 1 \quad and \quad s_{j}(t) = 1 \quad and \quad s_{h}(t) = 0 \quad and \quad otherwise,}
\end{array} \right.
\label{pcpd_generate_machanism}
\end{equation}
\end{figure*}

Like many non-equilibrium models, the PCPD is easy to simulate on computers, but hard to implement with experiments. Moreover, the exact analytical solution of the PCPD is not yet available. It does not have rapid time inversion symmetry. Analogous to many non-equilibrium models, the PCPD is described by four independent critical exponents ($\beta$, $\beta^{'}$, $\nu_{\perp}$, and $\nu_{\parallel}$). However, its universality class category has become one of the unsolved and controversial problems of non-equilibrium critical phenomena \cite{henkel2008non,henkel2004non}, which has received considerable amount of attention.

Take (1+1)-dimensional lattice as an example, the PCP has infinite number of absorbing states, whereas the PCPD can only have at most two absorbing states, including the one with all empty sites and the one with a single diffusion particle. The particle annihilation of the PCPD at absorbing phases shows an algebraic decay vs time, which is a piece of evidence that the PCPD may not belong to the DP universality class. In Monte Carlo simulations \cite{odor2003critical,deng2020coupled,odor2000critical}, usually we set the sum of diffusion, annihilation and proliferation probability of a single particle in the PCPD to be $1$. In addition, numerical simulations indicate that the upper critical dimension of the PCPD is $d_{c} = 2$, and that of the DP is $d_{c} = 4$. It is worth noting that for numerical results, some variables (such as particle density and pair density) do not obey exact power-law distributions. This could be one of the factors that $p_{c}$ and critical exponents of the PCPD are dependent on diffusion rates. In general, the critical behaviors of the PCPD still need to be unveiled, with higher measurement accuracy, which is exactly one of the motivations of this work.

\subsection*{Methods of unsupervised ML}
\subsubsection*{PCA}
As one of the most commonly used dimensionality reduction \cite{jolliffe2016principal,cichocki2009nonnegative,vesselinov2019unsupervised} and visualization techniques, PCA \cite{jolliffe2016principal} transforms a set of potentially linearly correlated data into a set of linearly unrelated variables. The first principal component is the data with the highest variance after transformation, and the second principal component is the data with the second-highest variance, and so on and so forth. There are two feasible methods to achieve PCA dimensionality reduction. The first one is on the strength of the eigenvalue decomposition of the covariance matrix, and the second one is on the strength of the singular value decomposition (SVD) of the original matrix. These two methods are intrinsically related, and please refer to Appendix A for details.

\subsubsection*{Autoencoder}
PCA is a linear dimensionality reduction algorithm, which eliminates redundancy by reducing the spatial dimension, and uses fewer features to describe data information as completely as possible. Autoencoder has advantages in feature extraction of linear data, nonlinear data denoising, image recognition, image compression, visual dimensionality reduction, and feature learning. In this paper, we focus on the latter two points.

We can refer to article \cite{tschannen2018recent} for recent advances of autoencoders, and \cite{dong2018review,zhai2018autoencoder} for systematic reviews of autoencoder and its variants. For example, sparse autoencoder is usually used to implement classification tasks, while denoising autoencoder can extract the most important features of data and learn their robust representation. Variational autoencoder is a generation system, which has similar functions as generative adversarial networks(GAN). Fully connected autoencoder generally ignores the spatial structure of the image. In this paper, we used the convolutional autoencoder \cite{masci2011stacked}. Please refer to Appendix B for an introduction to the principle including the loss function of autoencoder. 

To prevent over-fitting \cite{glassner2018deep}, we add the L2-norm ($\lambda/(2N) \sum\limits_{i} w^{2}_{i}$) to the loss function. The AdamOptimizer is used to speed up our neural networks. Our ML is implemented based on TensorFlow 1.15.

\subsection*{Data Sets and Results}
Before building an adaptable model, we first need to explicitly define an ML problem \cite{jung2022machine}. It includes data point selection, its features and labels, hypothesis space, and loss function. For (1+1)-dimensional PCPD, its data points can be regarded as a configuration with all time steps being generated from a single annihilation probability, as shown in Fig. \ref{pcpdshow}. It is characterized by the number of lattice sites contained in the configuration, including "1" (occupied site) and "0" (empty site). In unsupervised learning, including PCA and autoencoder studies, we do not need to label the configuration, but instead extract some information directly from the original data. In supervised learning, however, we have to label the configuration generated by MC simulations, only in this way can we train a special and reasonable model or hypothesis space. In the neural networks of this paper, our loss function is the mean cross-entropy between the label vector $y_{l}$ and the output layer vector $y$. In autoencoder neural networks, the label can be regarded as the original configuration itself. The main task of unsupervised learning's applying to phase transitions is to reduce the dimensionality of original high-dimensional data to extract critical points. In this paper, the main purpose of supervised learning is to make a binary classification for (1+1)-dimensional PCPD configurations, which are percolating and non-percolating phases. Depending on the binary classification output, we can also calculate some critical exponents by rescaling of the data collapse. Obtaining critical points and critical exponents is the fundamental motivation for us to study phase transition problems.

As explained in \cite{shen2021supervised}, for (1+1)-dimensional bond DP, the paper chooses the generated configuration with a fully occupied lattice as the initial condition. And this paper will deal with the data form in the same way. We employ Monte Carlo simulations to generate the raw input data(see Supplementary Section "PCPD's data sets for training and test") needed for ML. It is clear that for non-equilibrium lattice models, the characteristic time $t_{c} \sim L^{z}$ is much larger than the lattice size because of the extra time dimension. Therefore, for the sake of calculation, we will truncate the original data for ML. For example, in many cases, we will take $T=L$. This operation can reduce the amount of computation and the impact of fluctuations in some cases. Of course, such processing does not lower the accuracy of the results of ML and is therefore considered feasible.

In autoencoder, our configuration data is divided into a training set, a validation set, and a test set. We use the validation set to adjust the hyper-parameters so that the model is optimal.

\subsubsection*{Unsupervised ML of the PCP and the PCPD via PCA}
\begin{figure*}[ht]
\begin{tabular}{cc}
    \includegraphics[width=0.45\textwidth]{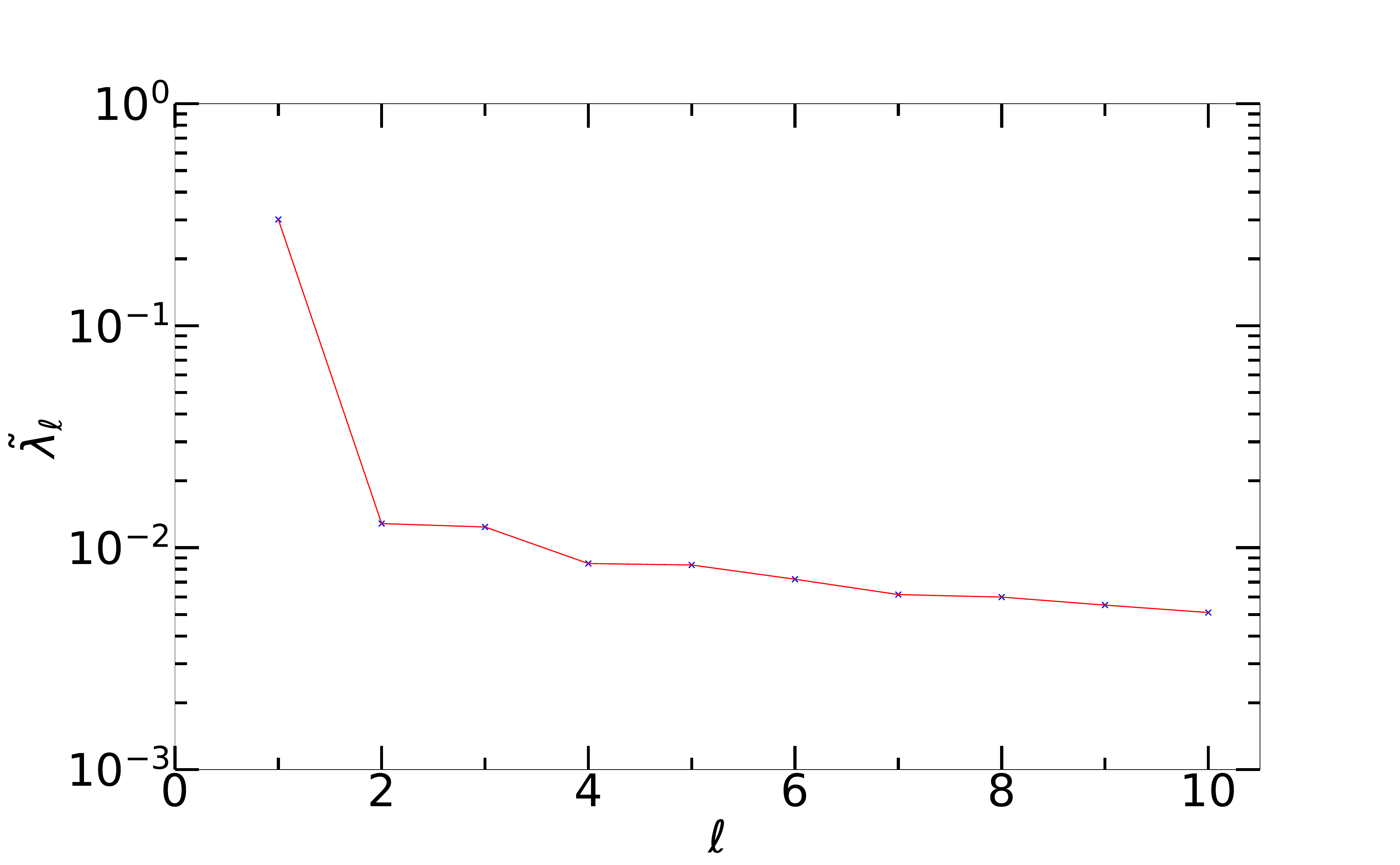} &
    \includegraphics[width=0.45\textwidth]{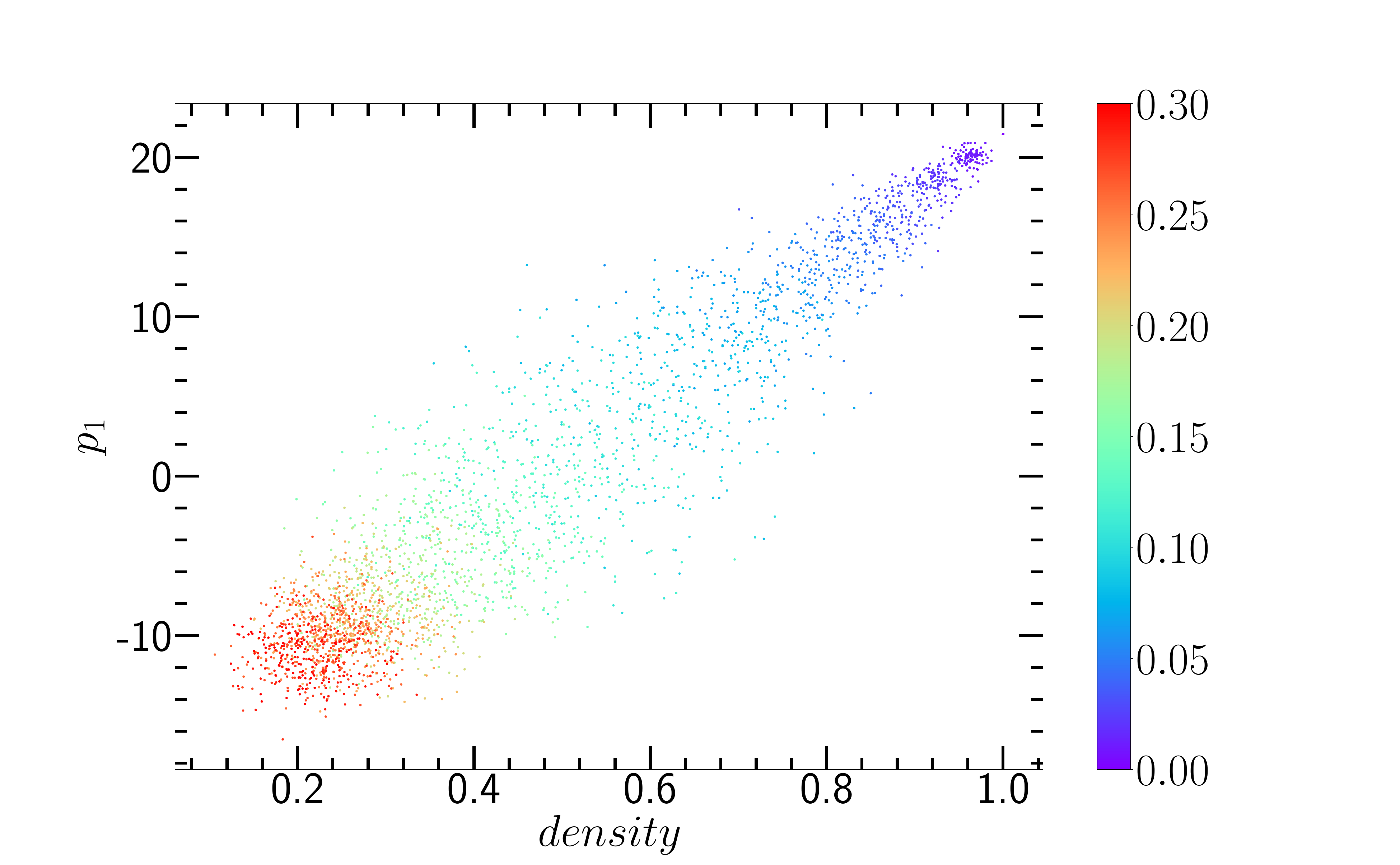} \\
    (a) &  (b)\\
    \includegraphics[width=0.45\textwidth]{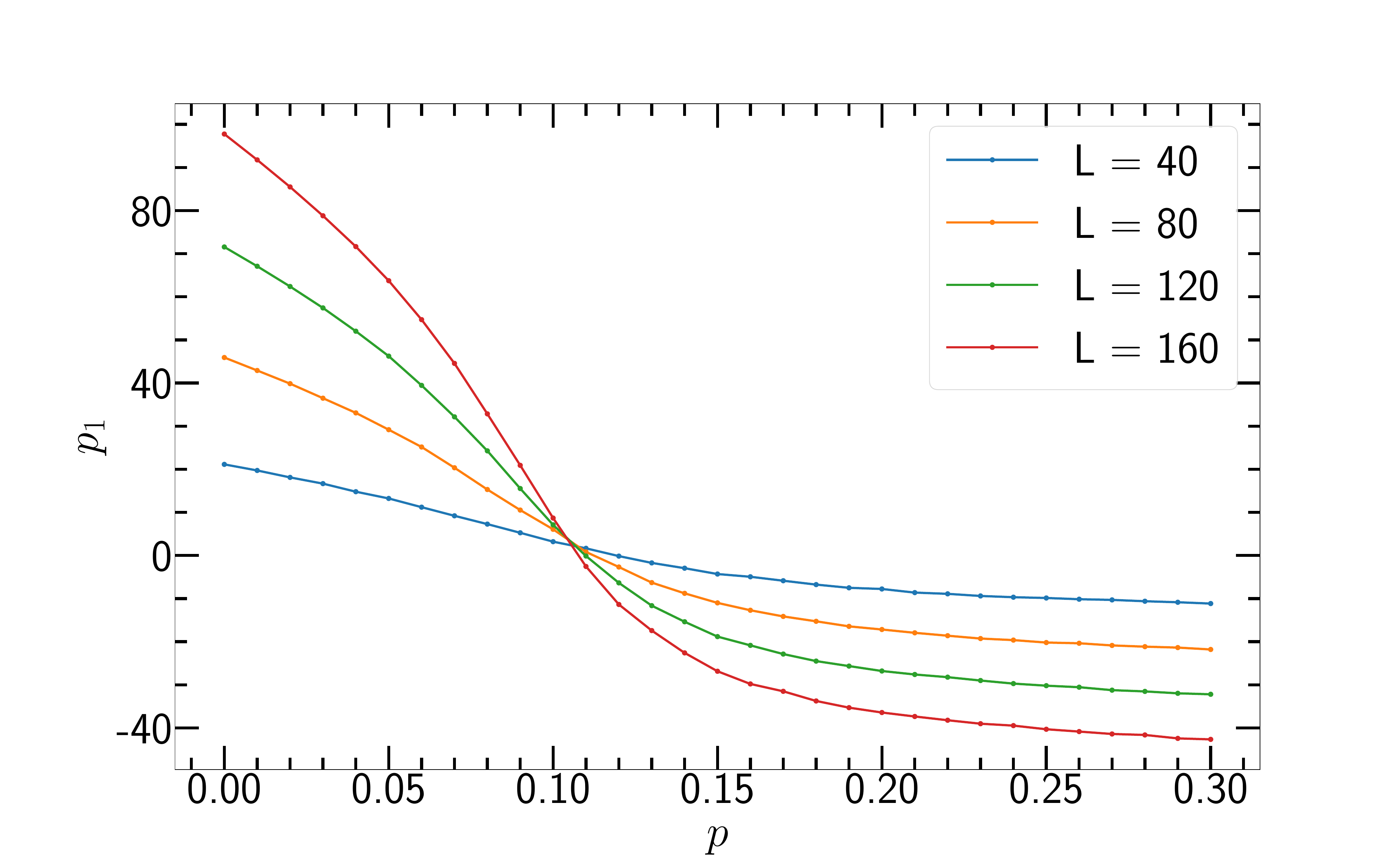} &
    \includegraphics[width=0.45\textwidth]{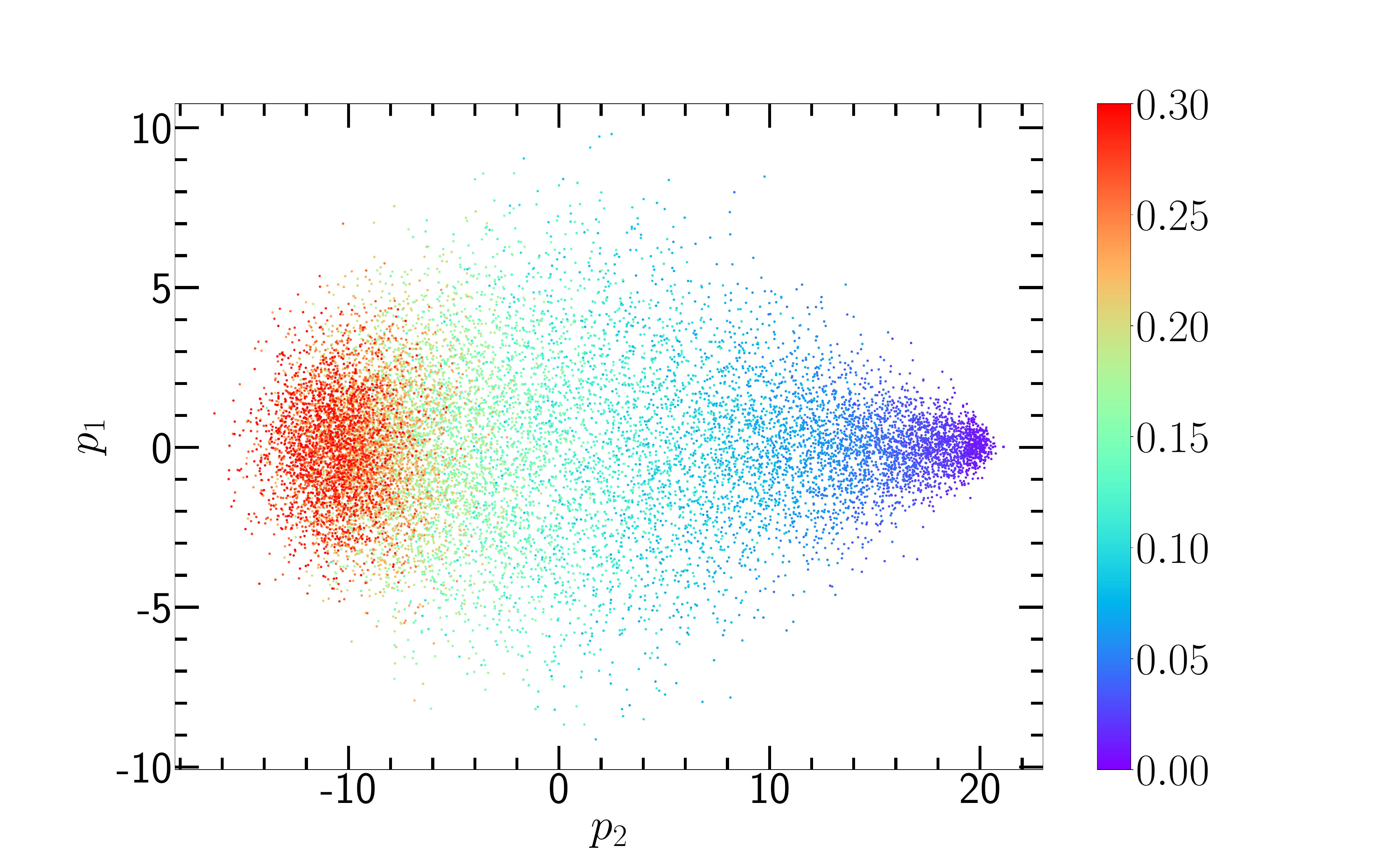} \\
    (c) & (d)
\end{tabular}
\caption{PCA results of (1+1)-dimensional PCPD, where $D=0.1$. \textbf{a}, The explained variance ratio $\widetilde{\lambda}_{\ell}$ from first ten principal components. \textbf{b}, $p_{1}$ versus $density$, each annihilation probability corresponding to $100$ samples.  \textbf{c}, $<p_{1}>$ ( ensemble averages of $1000$ runs) versus $p$. System sizes $L = 40, 80, 120, 160$ are represented by different colors, respectively. \textbf{d}, The clustering over $31$ annihilation probabilities.}
\label{pca_pcpd}
\end{figure*}

\begin{figure*}[!thb]
\begin{tabular}{cc}
    \includegraphics[width=0.45\textwidth]{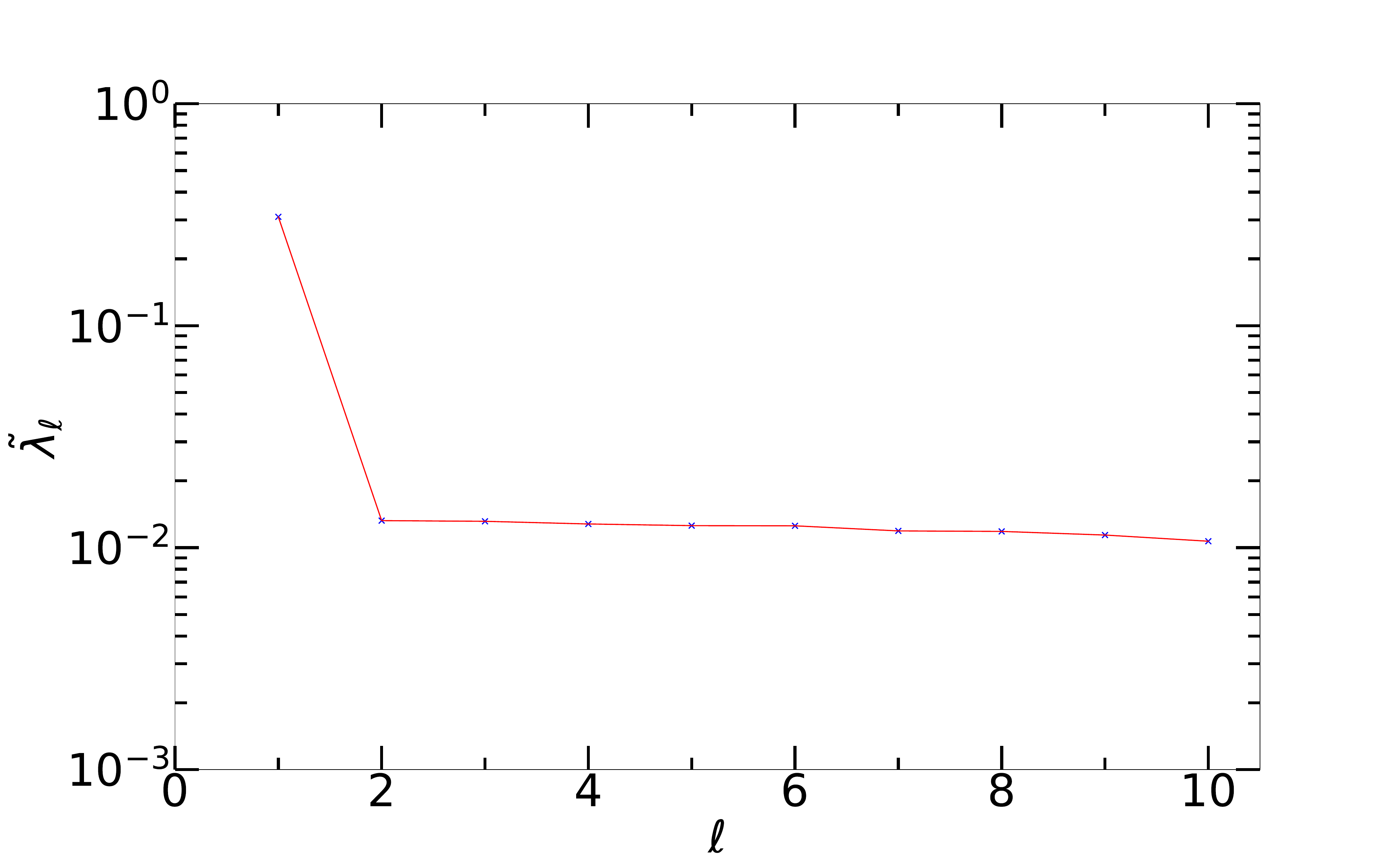} &
    \includegraphics[width=0.45\textwidth]{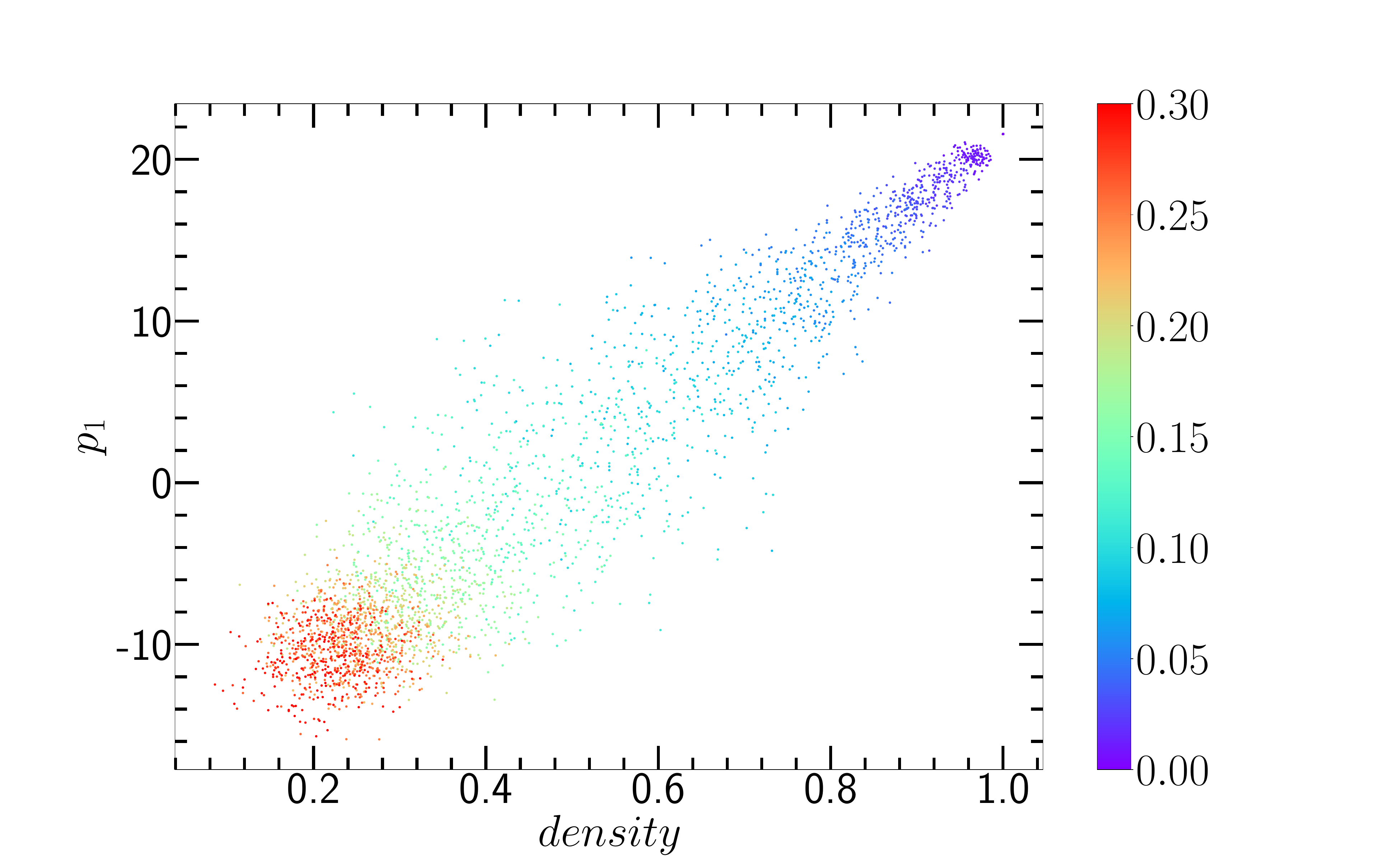} \\
    (a) &  (b)\\
    \includegraphics[width=0.45\textwidth]{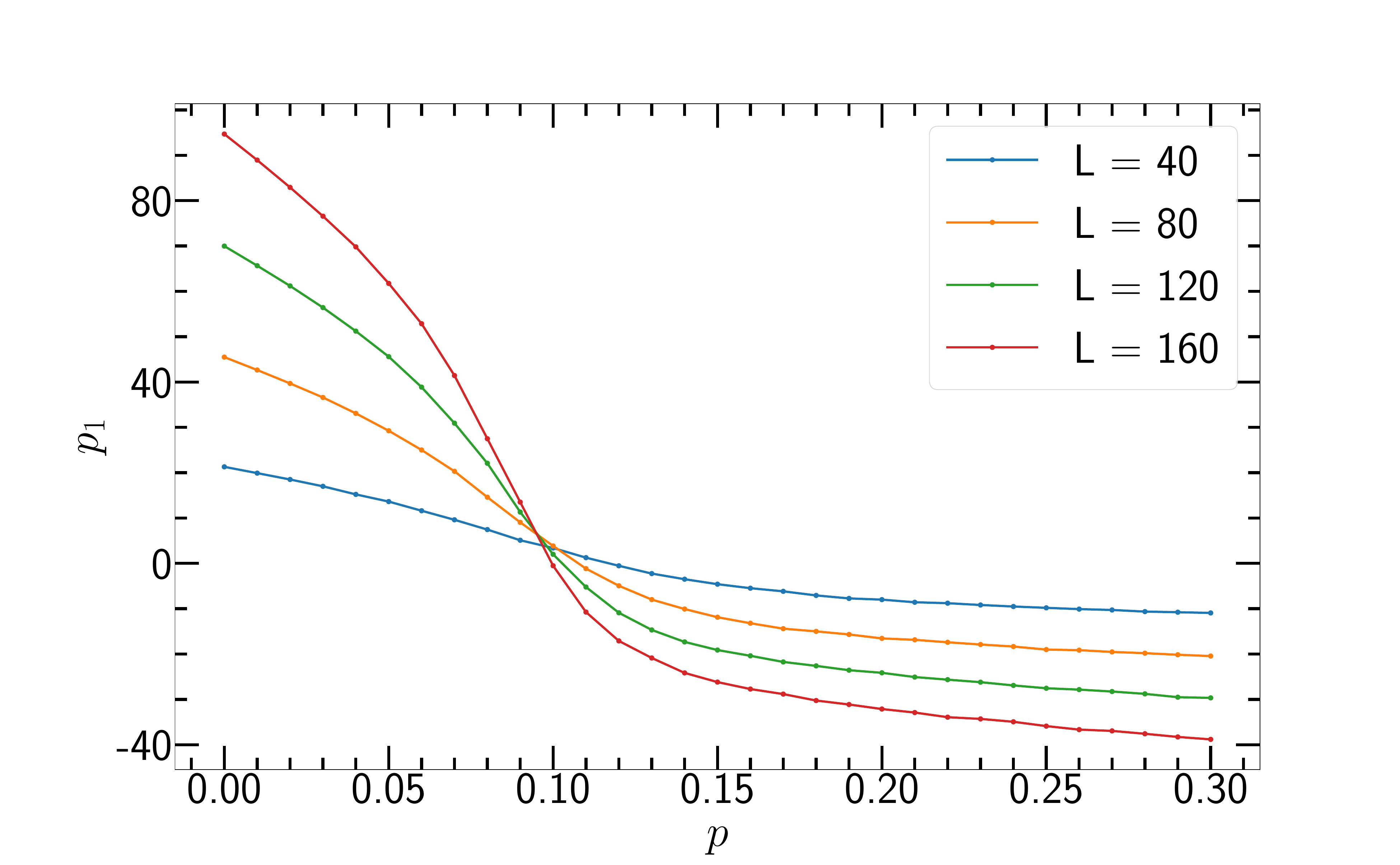} &
    \includegraphics[width=0.45\textwidth]{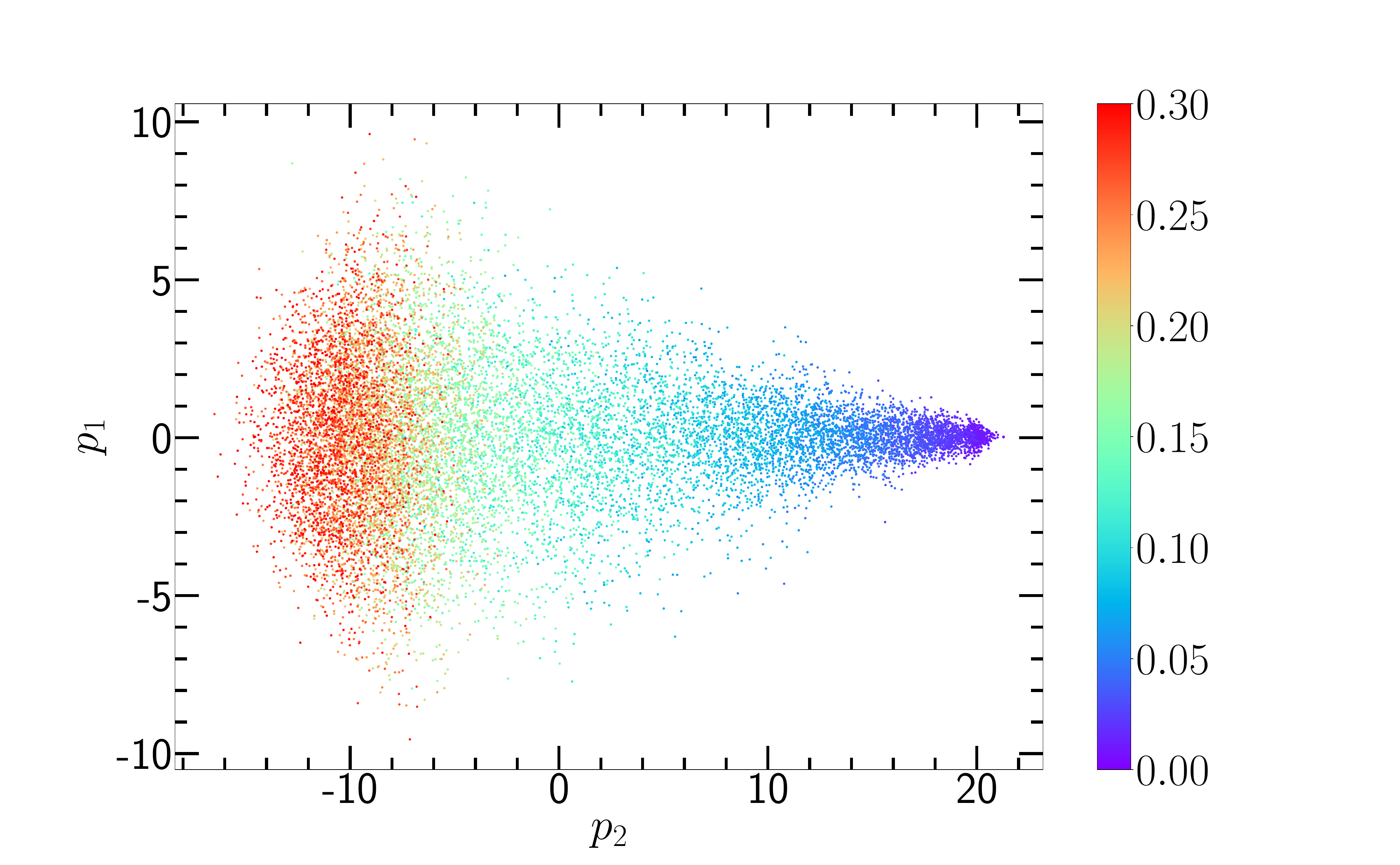} \\
    (c) & (d)
\end{tabular}
\caption{PCA results of (1+1)-dimensional PCP. \textbf{a}, The explained variance ratio $\widetilde{\lambda}_{\ell}$ from first ten principal components. \textbf{b}, $p_{1}$ versus $density$, each annihilation probability corresponding to $100$ samples.  \textbf{c}, $<p_{1}>$ (ensemble averages of 1000 runs ) versus $p$. System sizes $L = 40, 80, 120, 160$ are represented by different colors, respectively. \textbf{d}, The clustering over $31$ annihilation probabilities.}
\label{pca_pcp}
\end{figure*}

First, we perform PCA dimensionality reduction for (1+1)-dimensional PCPD. For this binary random reaction process, we learn the case of $D=0.1$. We use Monte Carlo simulations to generate configurations corresponding to different annihilation probabilities $p$. Here, we set the lattice size $L = 40$ and the time step $T = 40$. We select $31$ annihilation probabilities with an interval of $0.01$ between 0 to 0.3, and each of them generates $100$ samples. That is, the raw data matrix is $X_{3100 \times 1600}$. It is obvious that in Fig. \ref{pca_pcpd} (a), there is only one dominant principal component whose corresponding explained variance ratio is the largest. 

Fig. \ref{pca_pcpd} (b) presents the relationship between the first principal component and particle density. As can be seen from Fig. \ref{pca_pcpd} (b), the corresponding configurations of the same annihilation probability are arranged in close proximity. Simultaneously, the configurations within the range of $p = 0 \sim 0.3$ are approximately arranged in a straight line. We conclude
that the first principal component and particle density are proportional to one another. This means that such a correlation makes the particle density a physics quantity that can be related to the first principal component.

The relationship between the first principal component and the annihilation probability, is shown in Fig. \ref{pca_pcpd} (c). To achieve high accuracy, we generated $1000$ samples for each annihilation probability to perform an ensemble average for the first principal component. By observing Fig. \ref{pca_pcpd} (c), we can see that the jumping location is the transition point. Fig. \ref{pca_pcpd} (d) is the result of mapping the first principal component and the second principal component onto a plane. This approach also has a good clustering effect for (1+1)-dimensional PCPD.

The article \cite{mudunuru2021deep} illustrates that using more principal components can capture and retain the most crucial information from the original data. Actually, in this paper, for the purpose of clustering or phase classification (extracting critical points), only one or two principal components are needed.

In addition, we also implemented PCA dimensionality reduction and visualization for (1+1)-dimensional PCP, as shown in Fig. \ref{pca_pcp}. In this case, the particle does not diffuse. We can conclude that, although the particle diffusion rate has changed, the unsupervised learning method PCA can promptly make an excellent clustering representation of (1+1)-dimensional PCPD and capture the critical point.
\subsubsection*{Unsupervised ML of PCP and PCPD via autoencoder}
\begin{figure*}[!th]
\begin{tabular}{cc}
    \includegraphics[width=0.4\textwidth]{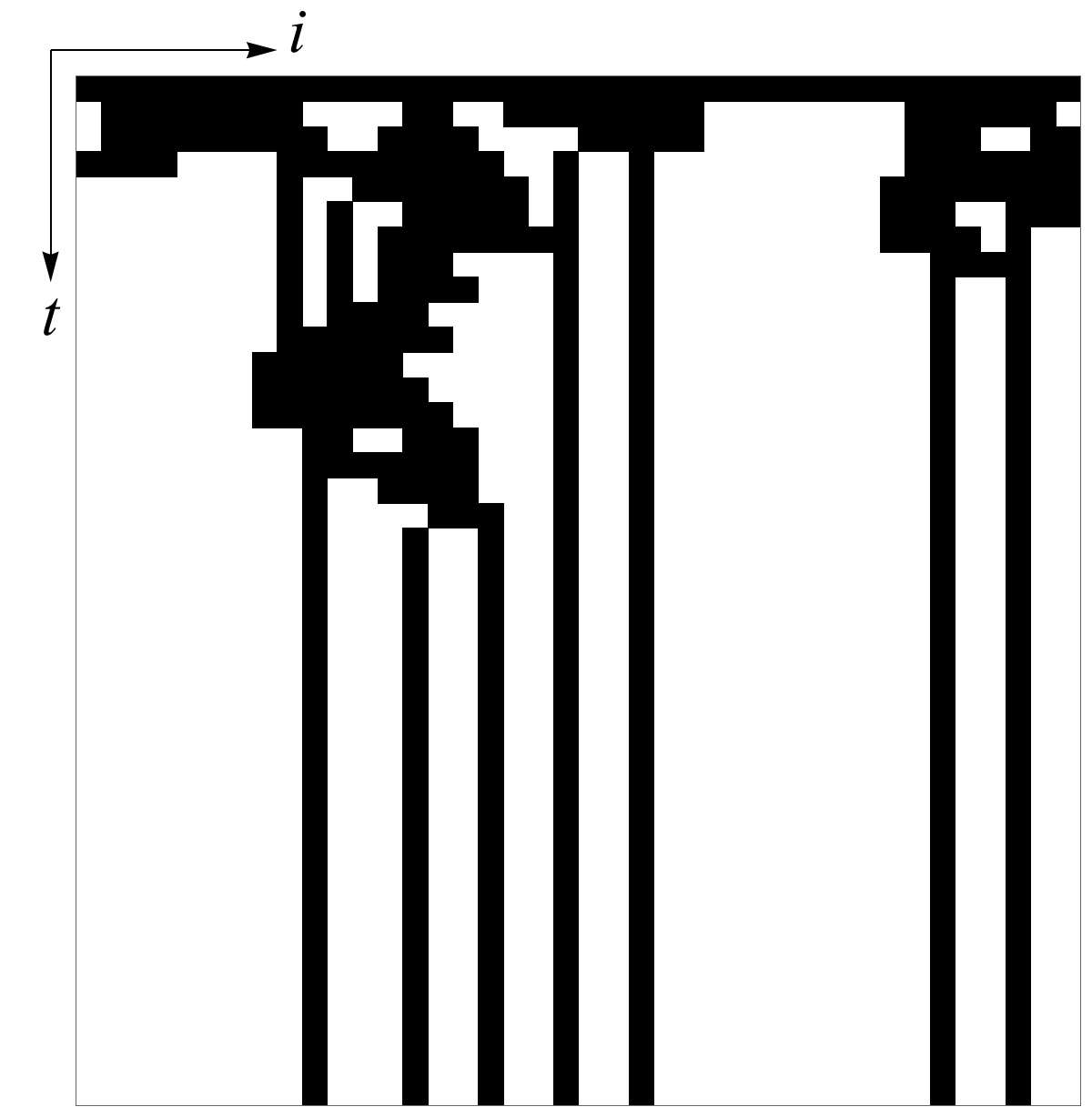} &
    \includegraphics[width=0.4\textwidth]{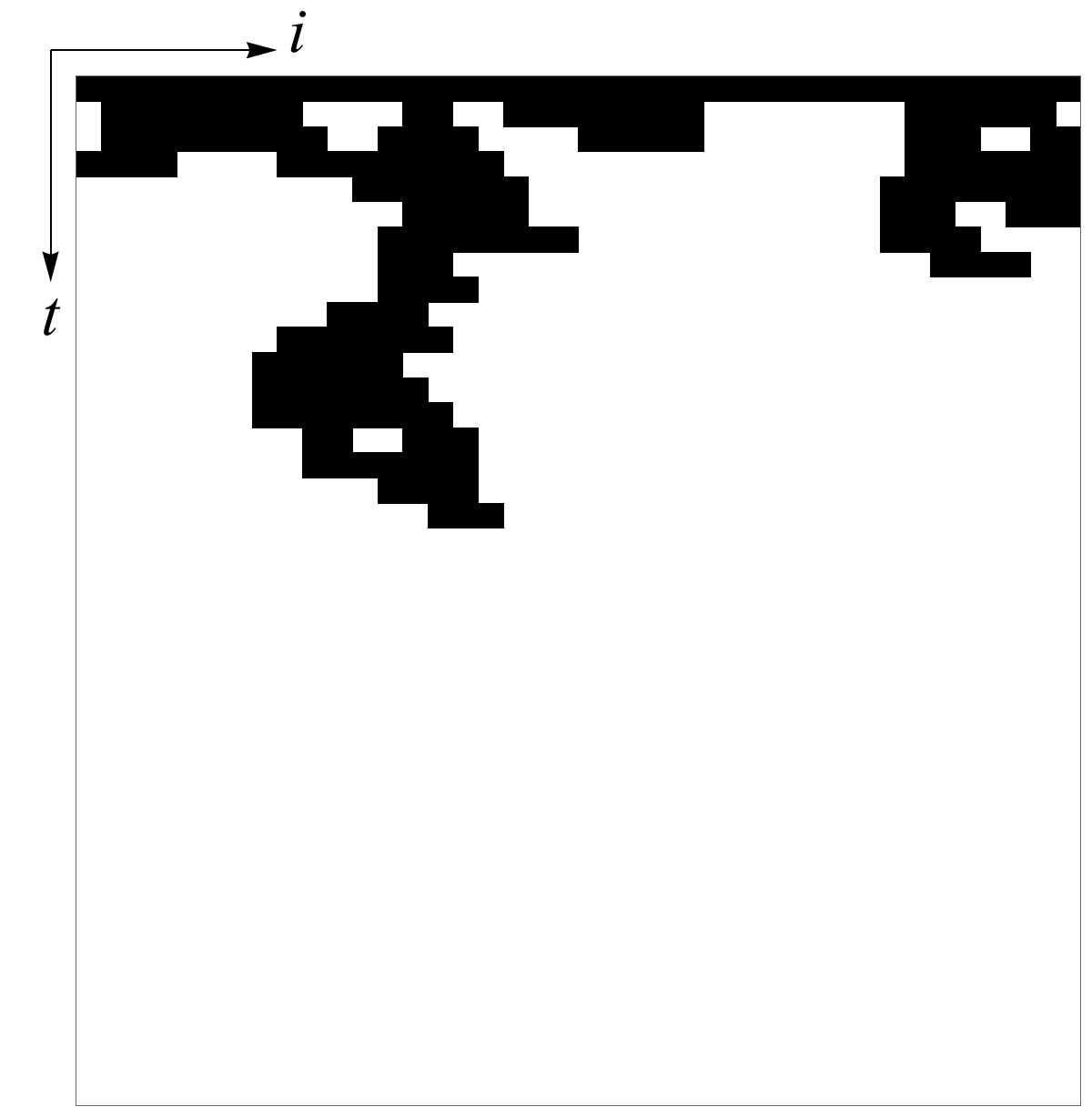}\\
    (a) &  (b) \\
    \includegraphics[width=0.45\textwidth]{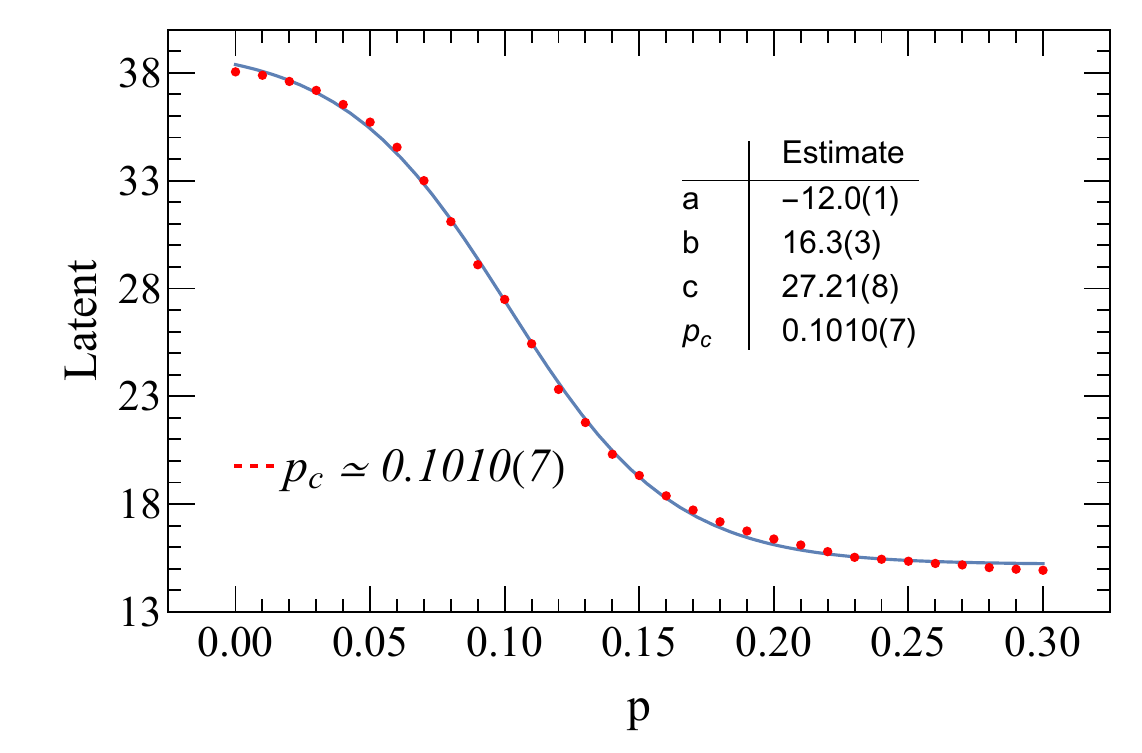} &
    \includegraphics[width=0.45\textwidth]{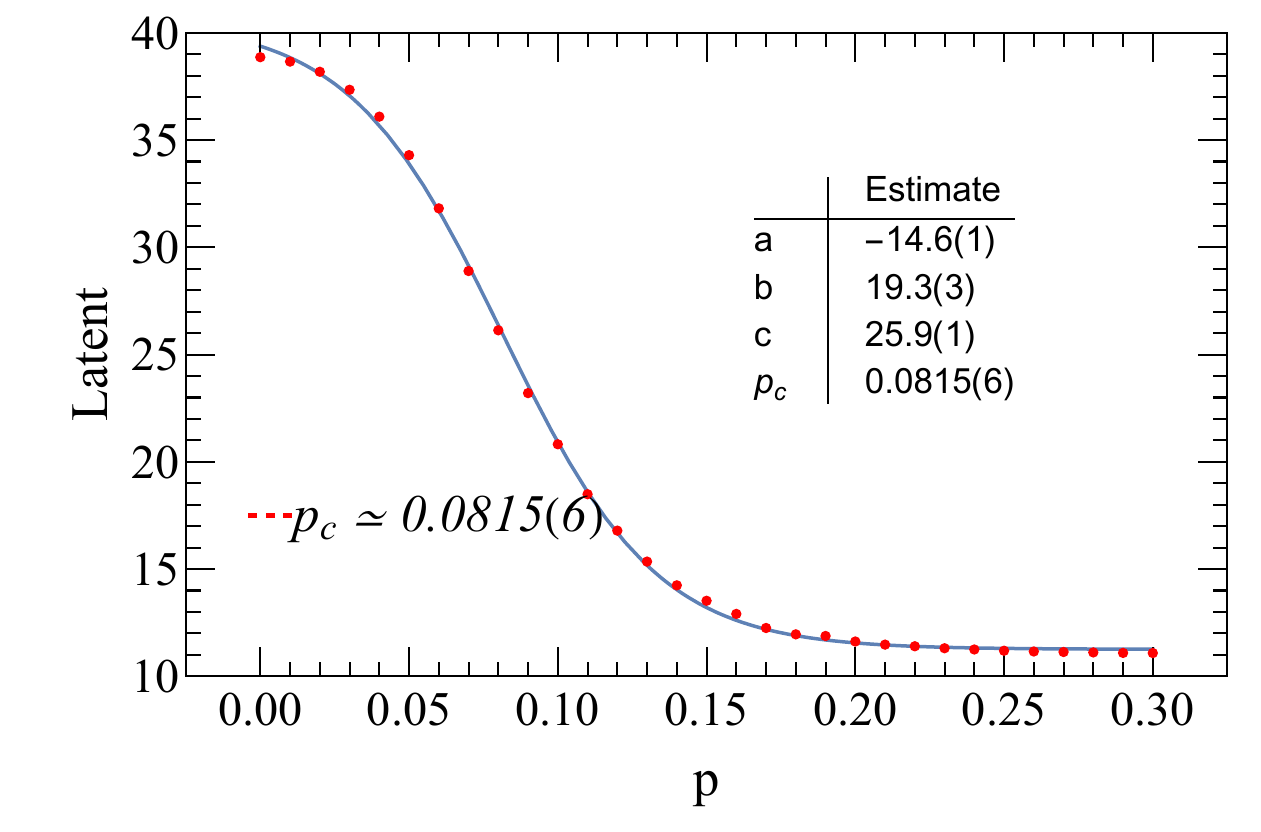} \\
    (c) & (d)
\end{tabular}
\caption{The top panels are configurations of (1+1)-dimensional PCP, where \textbf{a} is the raw configuration with lattice size $N=40$ and time step $t=40$. \textbf{b} represents the configuration of pair-particle of the lattice defined in \textbf{a}. \textbf{c} and \textbf{d}, autoencoder results of (1+1)-dimensional PCP. \textbf{c} gives the learning of configurations defined in \textbf{a}. \textbf{d}, gives the learning of configurations defined in \textbf{b}.}
\label{conf_ae_pcp_raw_pair}
\end{figure*}

Before employing the autoencoder learning of PCP, we need to pre-process the raw configurations. In phase transitions, the order parameter is the key quantity that separates the two different phases at the critical point. For PCP model, the density of pair particles, is one of two order parameters (the other one is density of single particles). Henceforth, feature engineering, for separating single and pair particles, is needed to deal with original configurations thereinbefore. Figs.\ref{conf_ae_pcp_raw_pair} (a) and (b) represent the raw configuration and the one where only pair particles are left, of the PCP model, respectively. After the preparations, we carry out autoencoder learning for these two configurations, and show the results at the bottom panels of Fig.\ref{conf_ae_pcp_raw_pair}, left and right. To determine the critical point, we perform a non-linear fitting in the form of the hyperbolic tangent function $a \times \tanh [b (p - p_{c})] + c$ (the blue lines of the bottom panels of Fig.\ref{conf_ae_pcp_raw_pair} are the fitting curves), and the location $p_{c}$ is the critical point we are looking for. The fitting technology we used is the NonlinearModelFit function of Mathematica, with which the Levenberg–Marquardt algorithm had been used with at least 10 iterations to minimize the sums of squares. We find that the predicted $p_{c} \simeq 0.0809$ of Fig.\ref{conf_ae_pcp_raw_pair} (d) is very close to that given by Monte Carlo simulations, where $p_{c} \simeq 0.0771$. Thus, we conclude that the pair-particle density can characterize the phase transition of the PCP in a good manner. The following learning results of the PCP are then all based on the configurations of pair particles only.

\begin{figure*}[!th]
\begin{tabular}{cc}
    \includegraphics[width=0.47\textwidth]{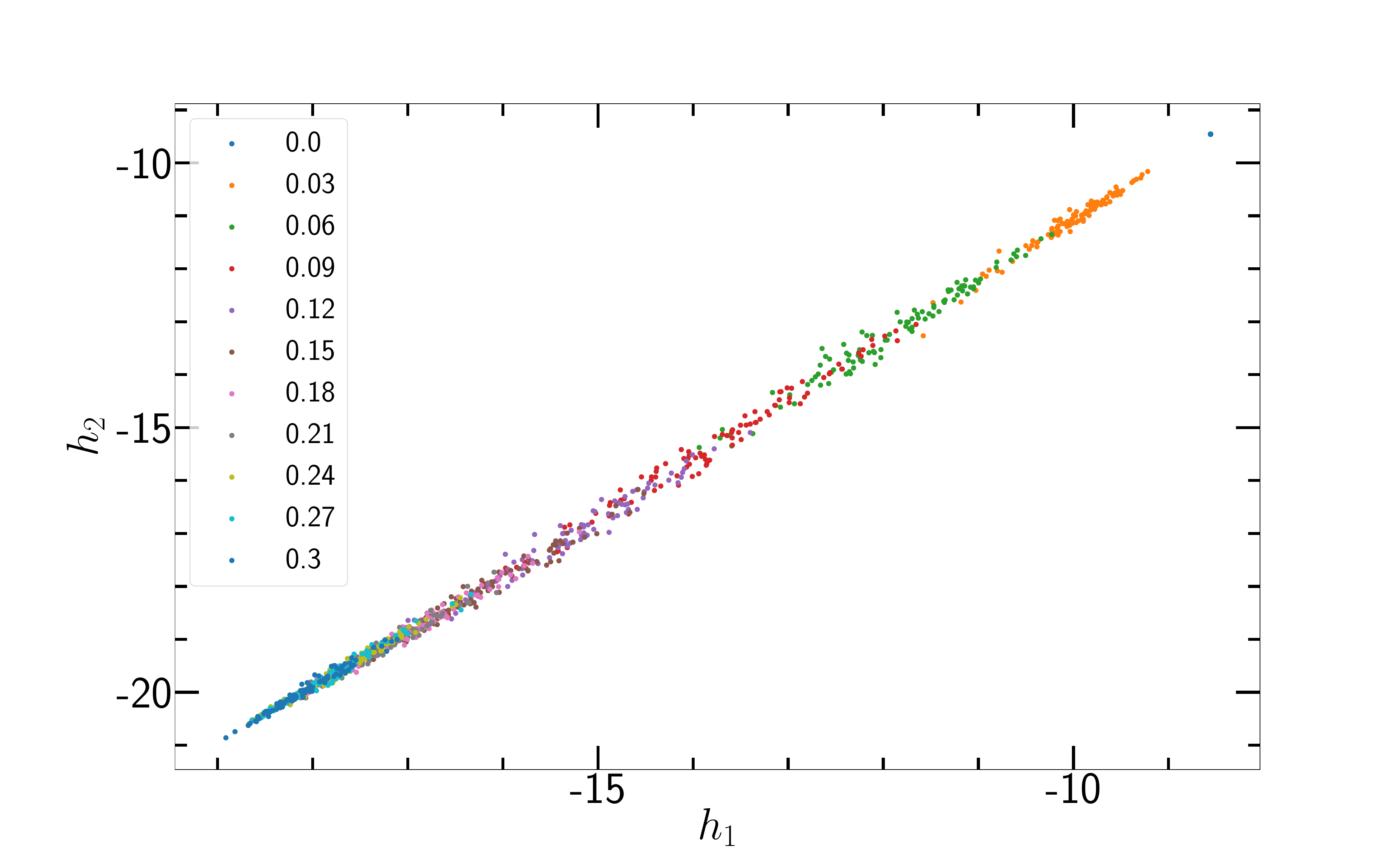} &
    \includegraphics[width=0.47\textwidth]{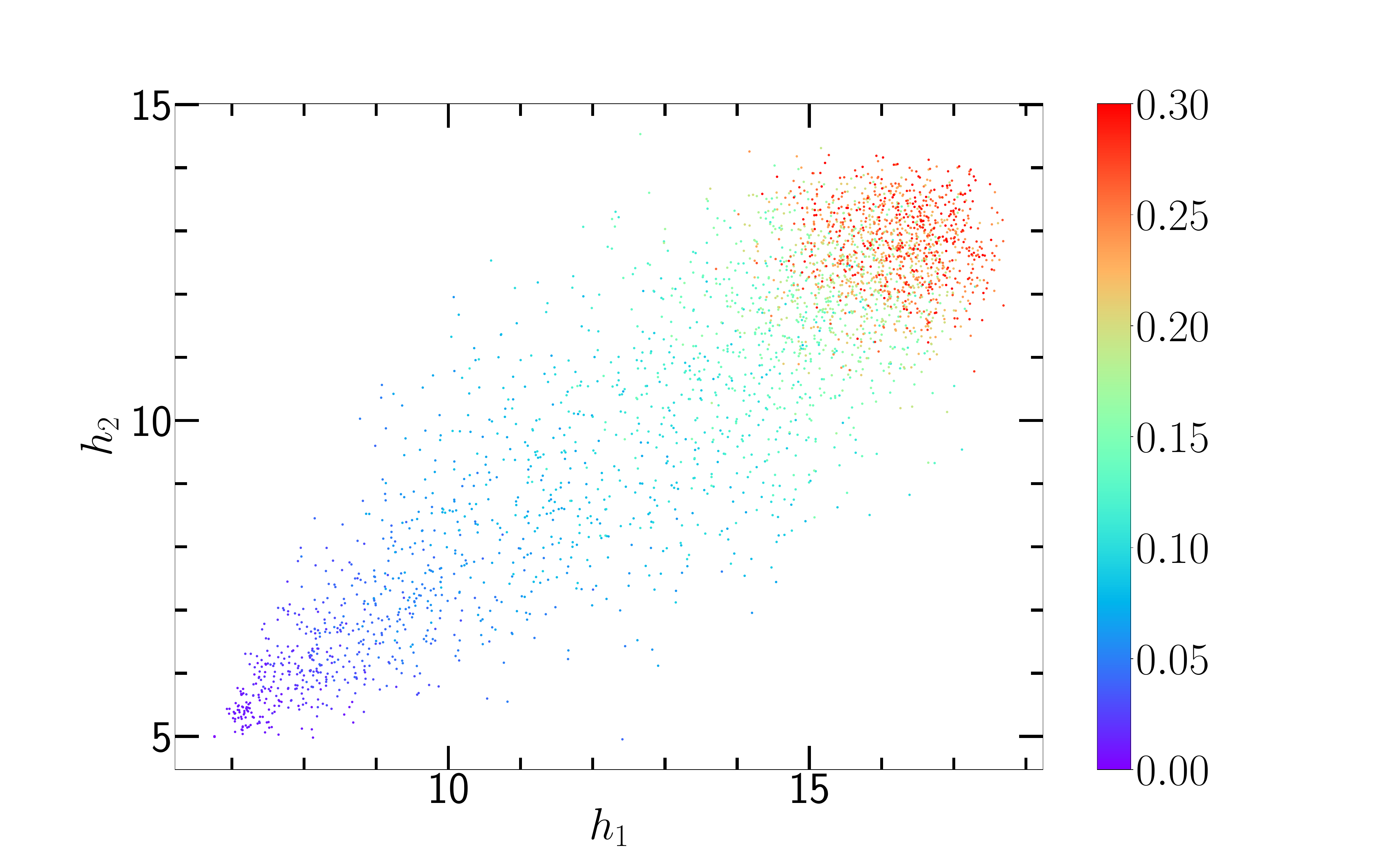} \\
    (a) &  (b) \\
    \includegraphics[width=0.45\textwidth]{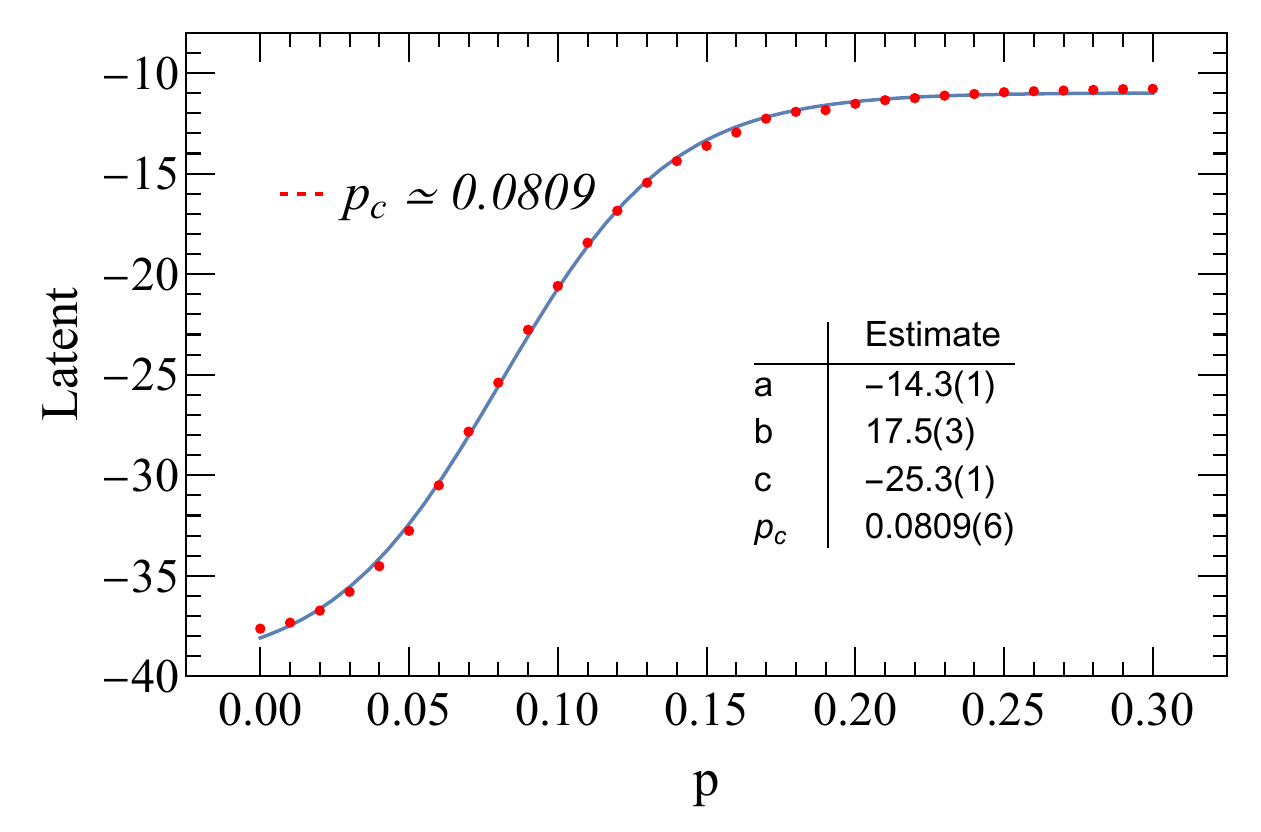} &
    \includegraphics[width=0.45\textwidth]{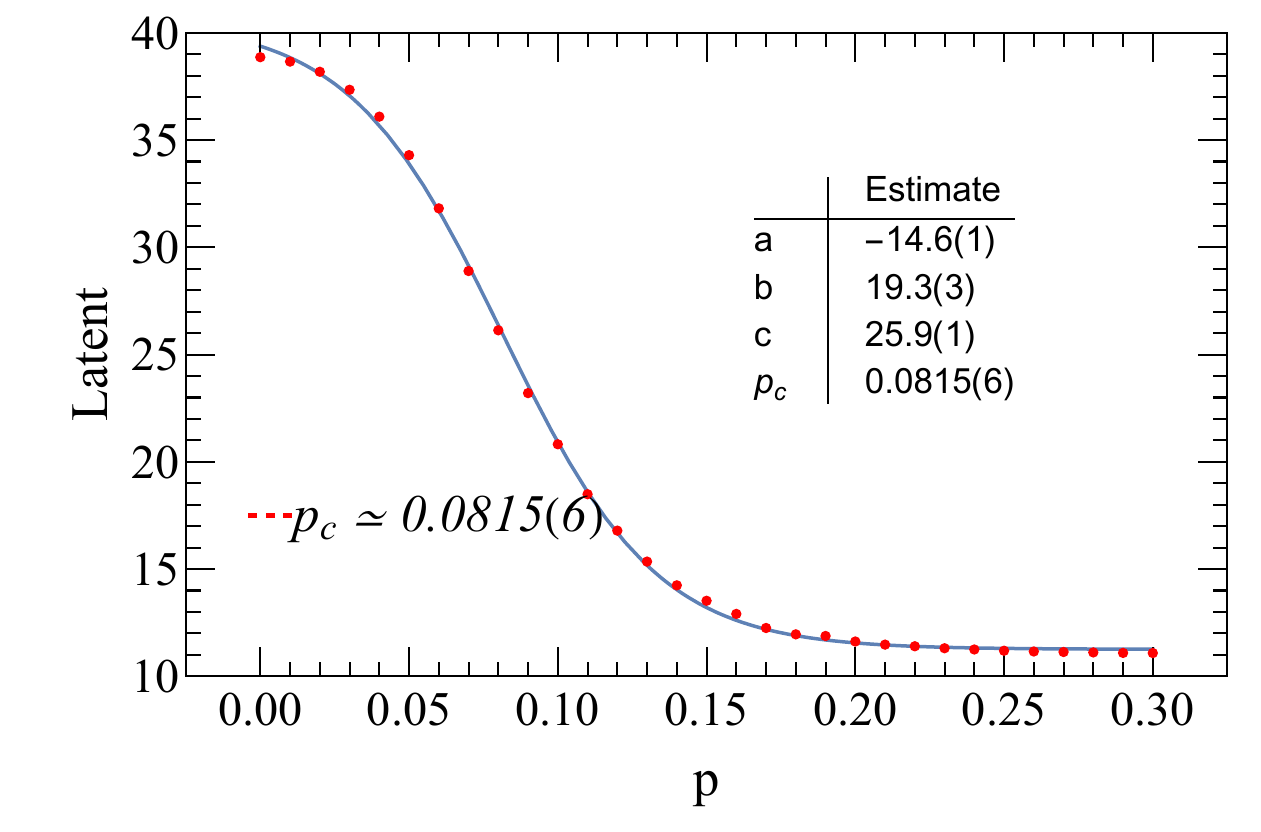} \\
    (c) & (d)
\end{tabular}
\caption{Autoencoder results of (1+1)-dimensional PCP. \textbf{a} and \textbf{b}, encoding of the raw PCP configurations onto the plane of the two hidden neuron activations $(h_{1}, h_{2})$. \textbf{c} and \textbf{d}, Encoding of the raw PCP configurations, using a single hidden neuron activation $Latent$ as a function of the annihilation probability. As seen, the critical threshold approximates the estimate by Monte Carlo simulations.}
\label{pcp_autoencoder}
\end{figure*}

Fig. \ref{pcp_autoencoder} shows the autoencoder learning results of (1+1)-dimensional PCP. The lattice size is $L=40$, and the time step is $T=40$. Figs. \ref{pcp_autoencoder} (a) and (b) represent two types of clustering, with configurations being generated by 11 and 31 different annihilation probabilities, respectively. As is known, the diffusion rate of PCP is $D=0$. Fig. \ref{pcp_autoencoder} (a) displays the clustering of the cases with $11$ different annihilation probabilities $p$, and for each $p$, $100$ samples of the configuration are generated. The number of potential neurons in the autoencoder network being set to 2, the eleven categories collapse onto a straight line. In Fig. \ref{pcp_autoencoder} (b), $31$ annihilation probabilities, between $0$ to $0.3$ and with an interval of $0.01$, are selected. Clustering results indicate that data points corresponding to smaller $p$ appear on the lower left (blue), while those corresponding to larger $p$  appear on the upper right (red). The middle segment is more scattered than the two ends, which by our speculation may belong to the critical regions.

In order to determine the critical probability $p_{c}$, the number of potential neurons of the convolutional autoencoder neural network is set to be $1$. After the neural network is trained, we got the latent variable shown in Figs. \ref{pcp_autoencoder} (c) and (d) where single potential variables are plotted as functions of $p$. The hyperbolic tangent function fitting yields $p_{c} = 0.081(1) $ and $p_{c} = 0.082(1)$ that are very close to the theoretical value of $p_{c} = 0.077092(1)$ \cite{henkel2008non}.

\begin{figure*}[htbp]
\begin{tabular}{ccc}
    \includegraphics[width=0.25\textwidth]{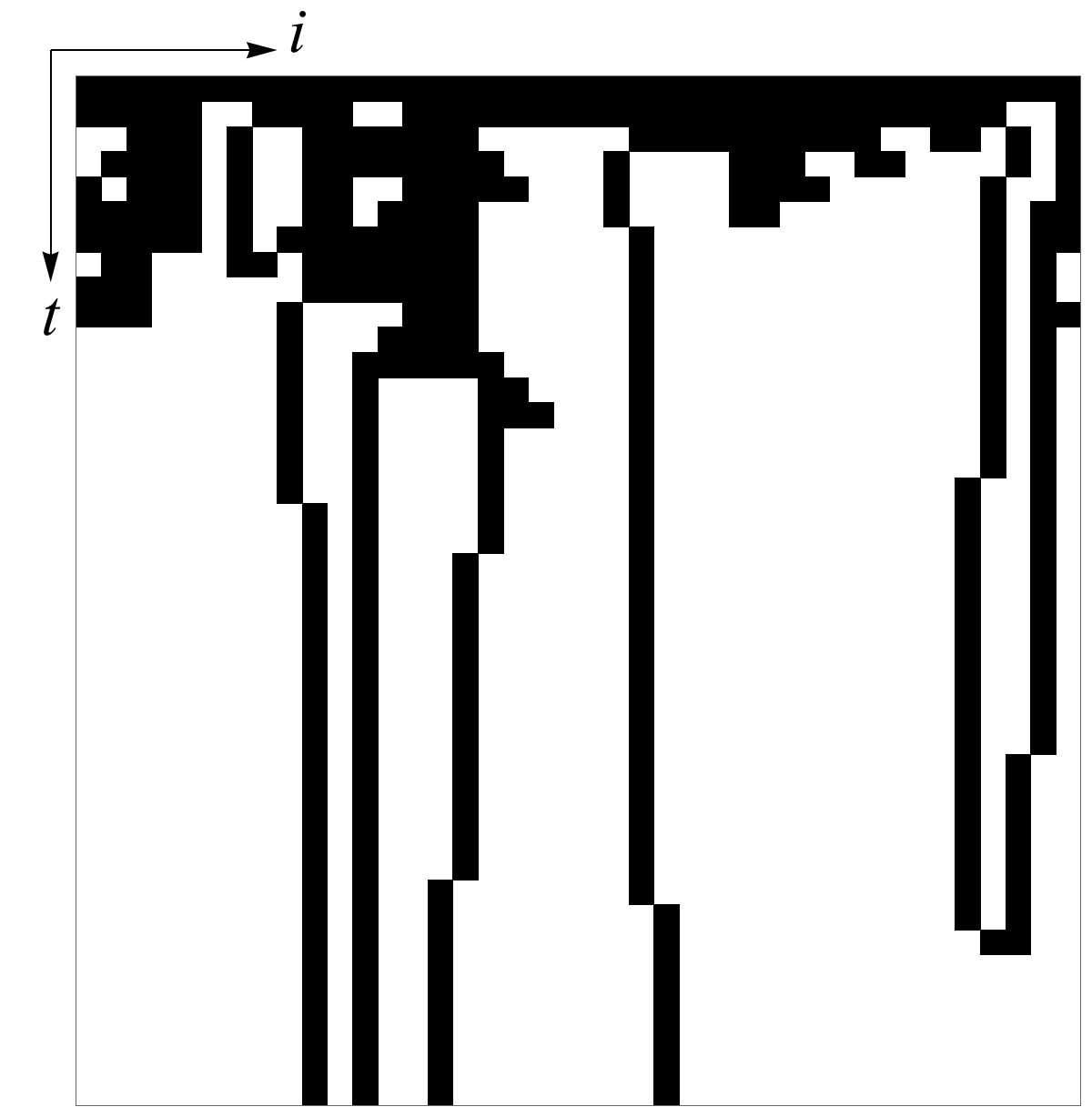} &
    \includegraphics[width=0.25\textwidth]{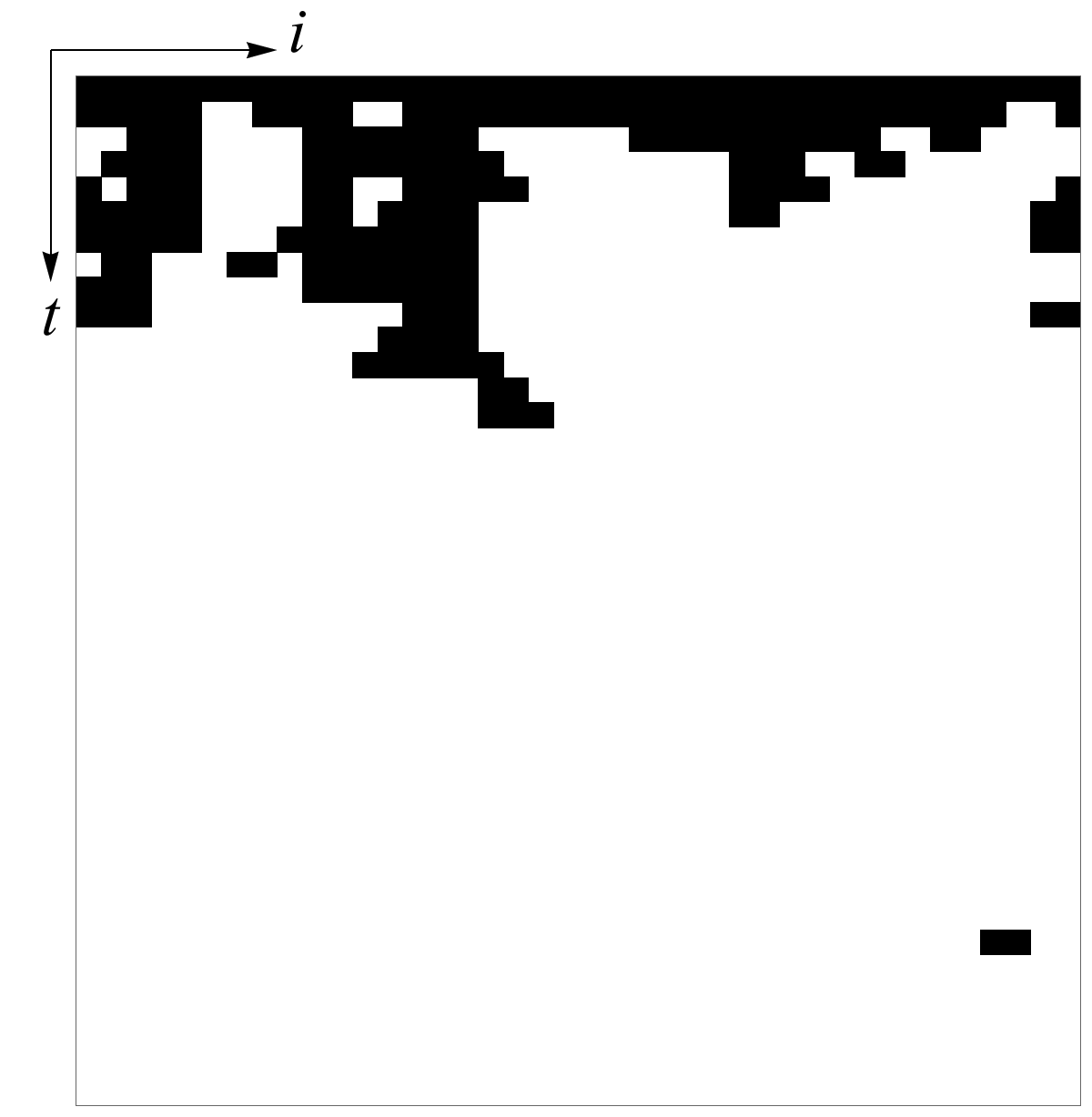}     &    \includegraphics[width=0.25\textwidth]{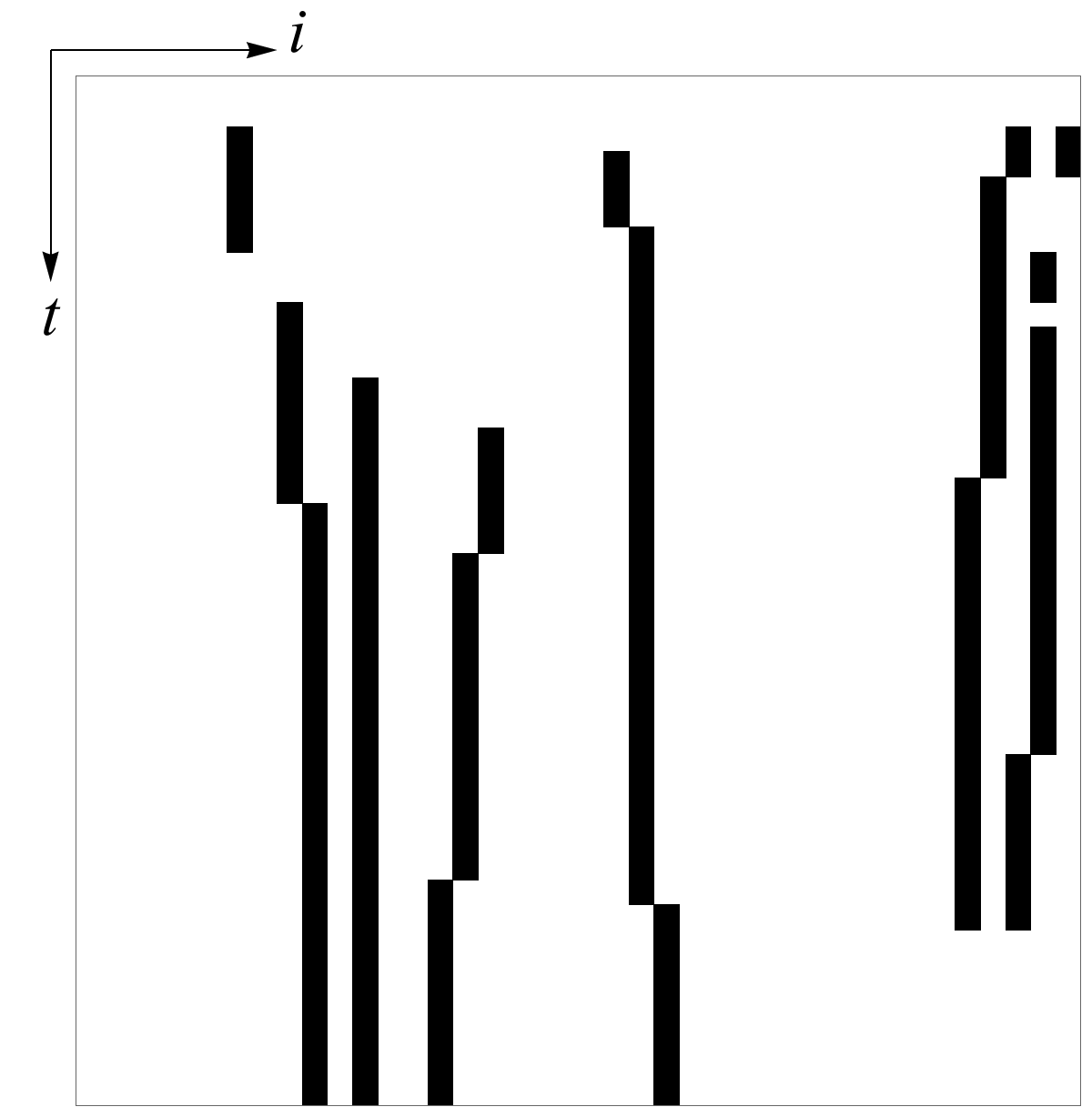}\\
    (a) &  (b) &  (c) \\
    \includegraphics[width=0.28\textwidth]{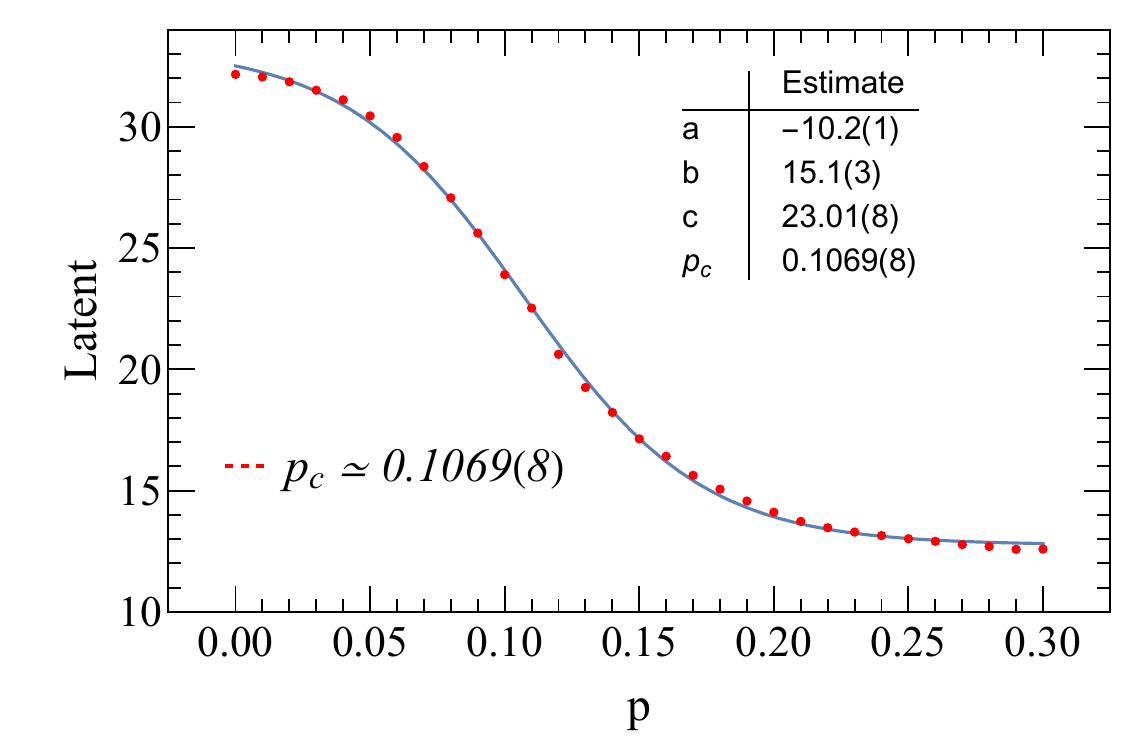} & \includegraphics[width=0.28\textwidth]{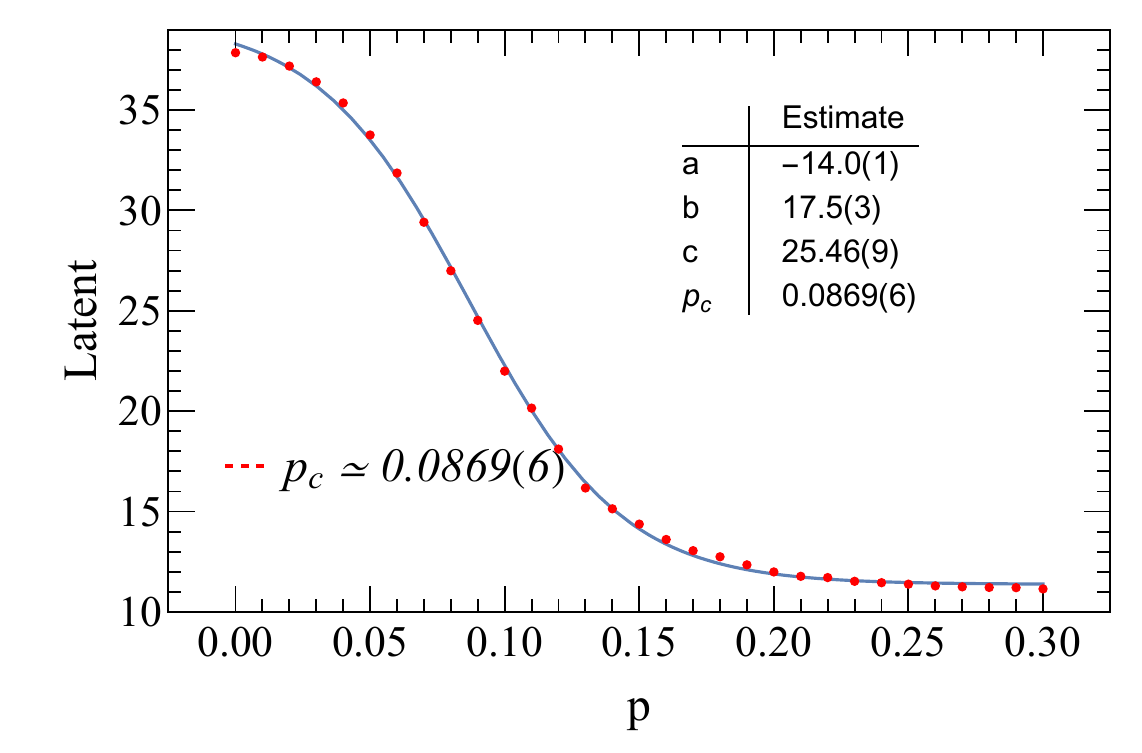} &
    \includegraphics[width=0.28\textwidth]{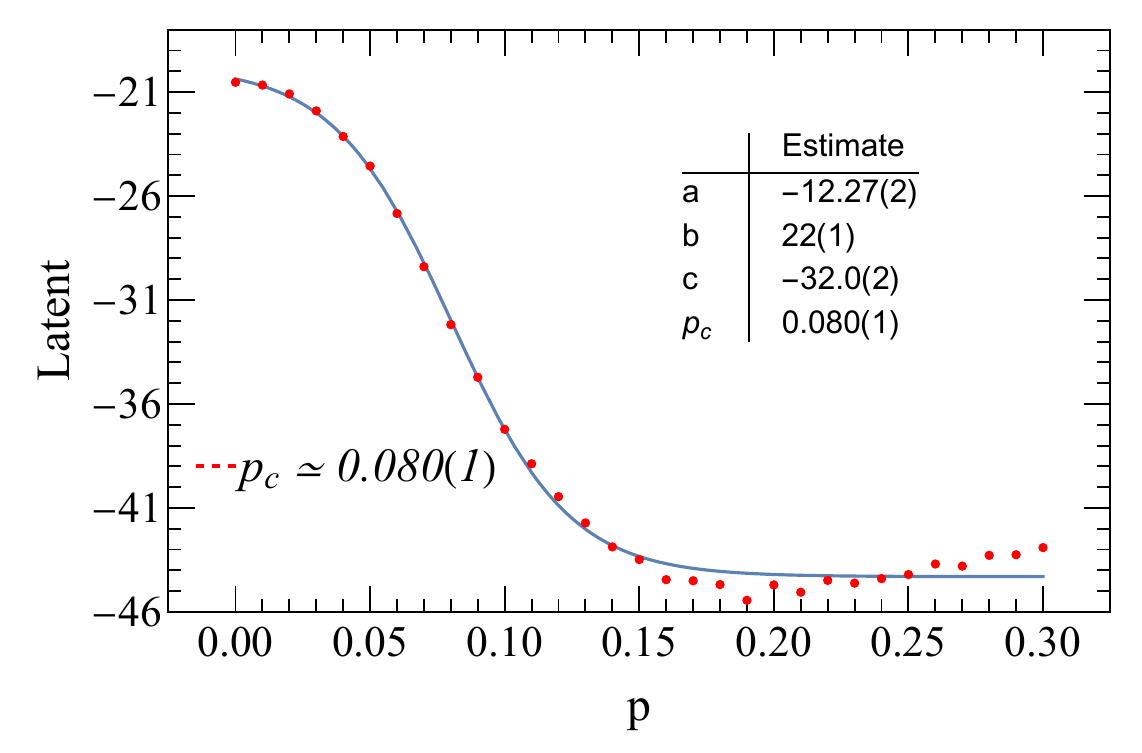} \\
    (d) & (e) & (f)
\end{tabular}
\caption{The top panels are configurations of (1+1)-dimensional PCPD, where diffusion rate $D=0.05$, lattice size $N=40$ and time step $t=40$. \textbf{a}-\textbf{c} represent raw, pair-particle and single-particle configurations, respectively. \textbf{d}-\textbf{f}, autoencoder learning results of (1+1)-dimensional PCPD. \textbf{d} corresponds the learning of raw configurations. \textbf{e}, the learning of pair-particle configurations. \textbf{f}, the learning of single-particle configurations.}
\label{conf_ae_pcpd_raw_pair_single}
\end{figure*}

Analogous to the PCP model, the configurations of the PCPD model for autoencoder learning are selectively chosen. Three different configurations of the PCPD are shown in the top panel of Fig.\ref{conf_ae_pcpd_raw_pair_single}, and their corresponding autoencoder learning results are given in the bottom panel of the same figure, respectively. It is found that only the learning based on the original configurations agrees well with the Monte Carlo simulations. Therefore, for the PCPD model, we choose the original configurations for autoencoder learning. Theoretically, the order parameter of the PCPD model is expressed by the sum of the densities of pair particles and single particles.

\begin{figure*}[htbp]
\begin{tabular}{ccc}
    \includegraphics[width=0.45\textwidth]{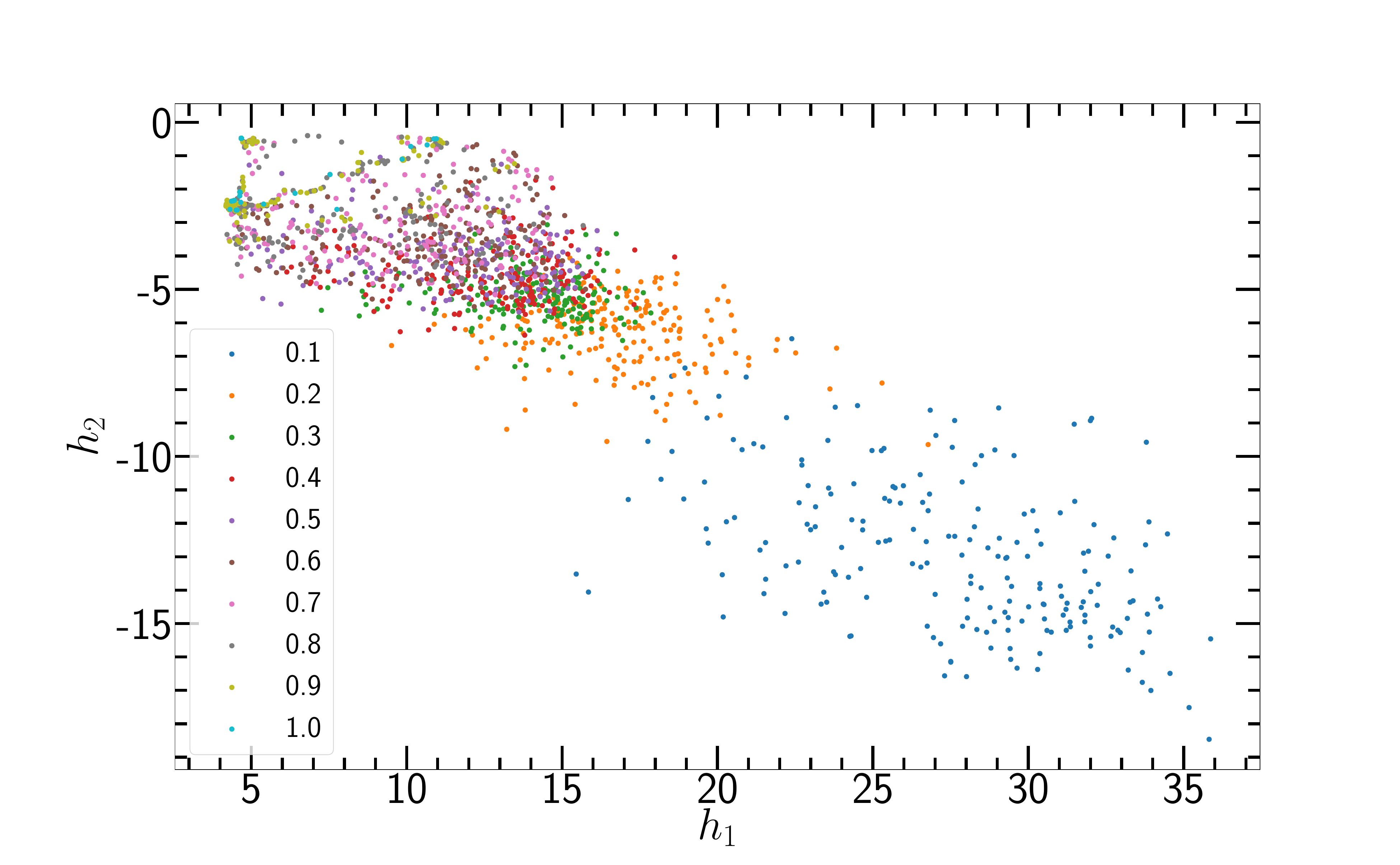} &
    \includegraphics[width=0.45\textwidth]{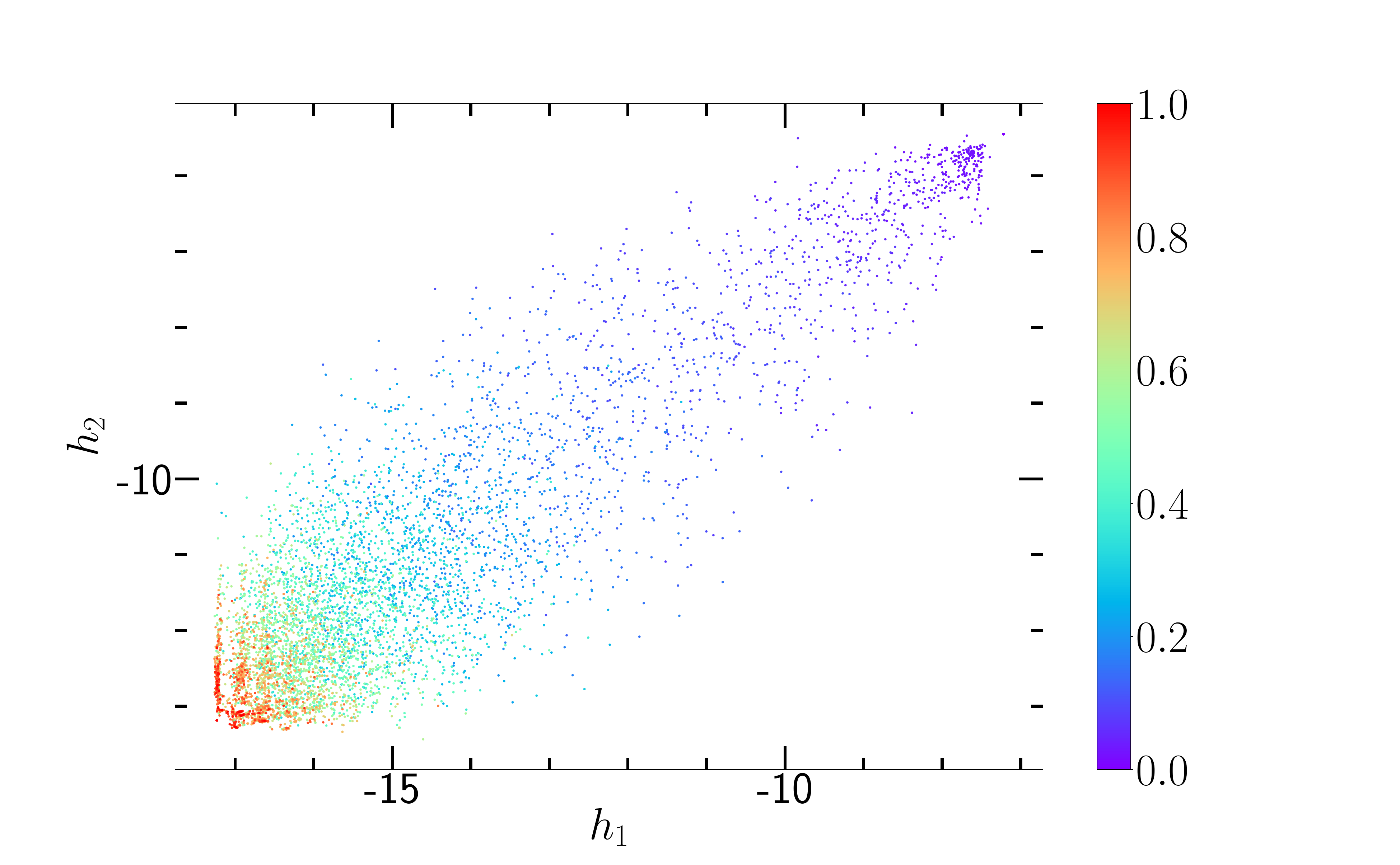} \\
      (a) &  (b)
\end{tabular}
\caption{Autoencoder learning results of (1+1)-dimensional PCPD, where $D = 0.05$.}
\label{d005_pcpd_p1p2}
\end{figure*}

\begin{figure*}[htbp]
\begin{tabular}{ccc}
    \includegraphics[width=0.28\textwidth]{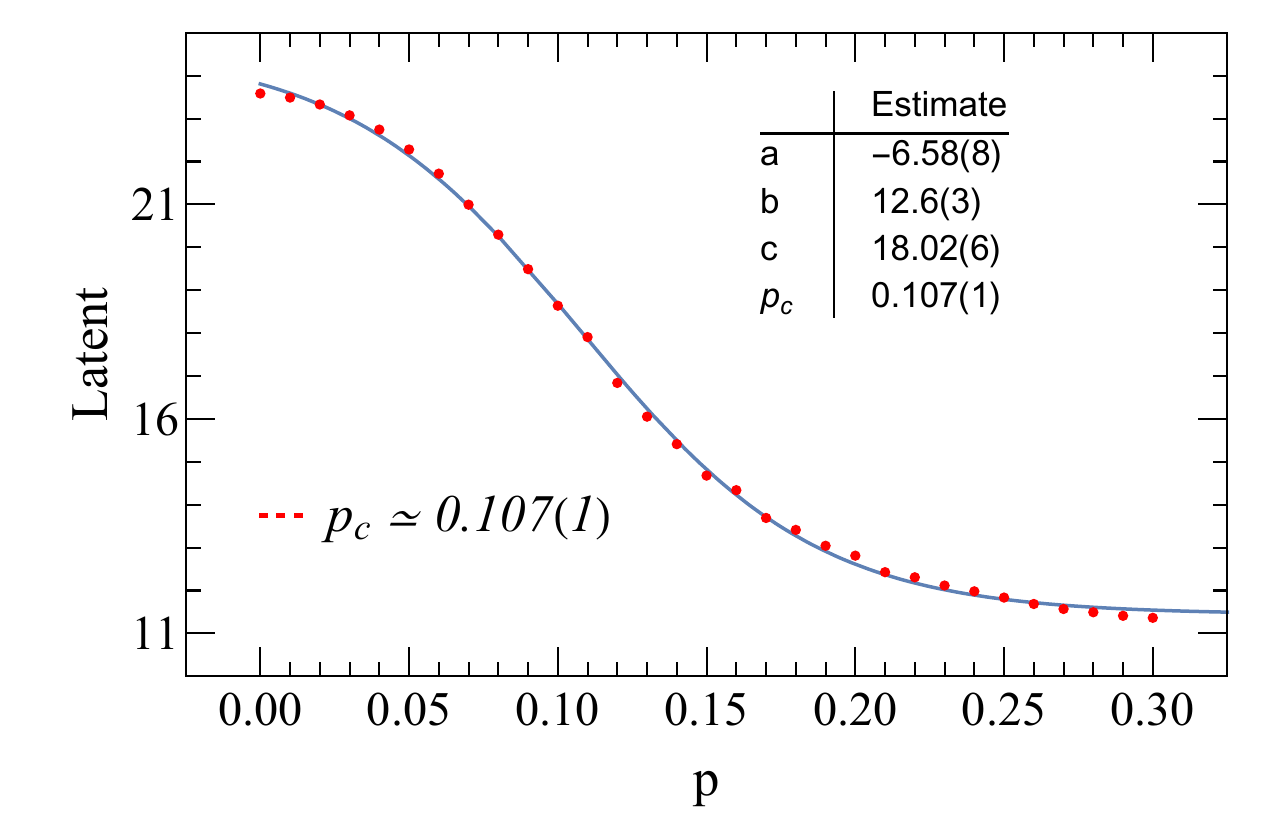} &
    \includegraphics[width=0.28\textwidth]{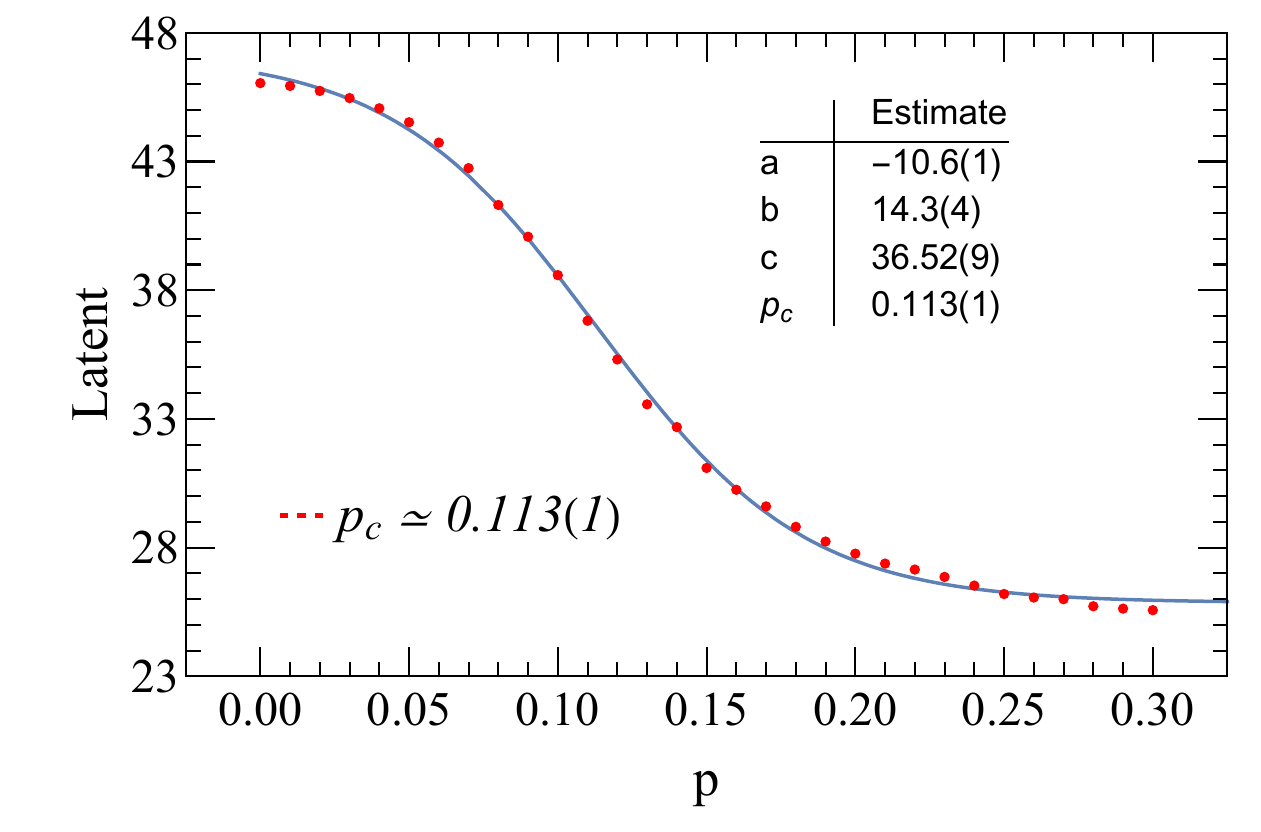}     &    \includegraphics[width=0.28\textwidth]{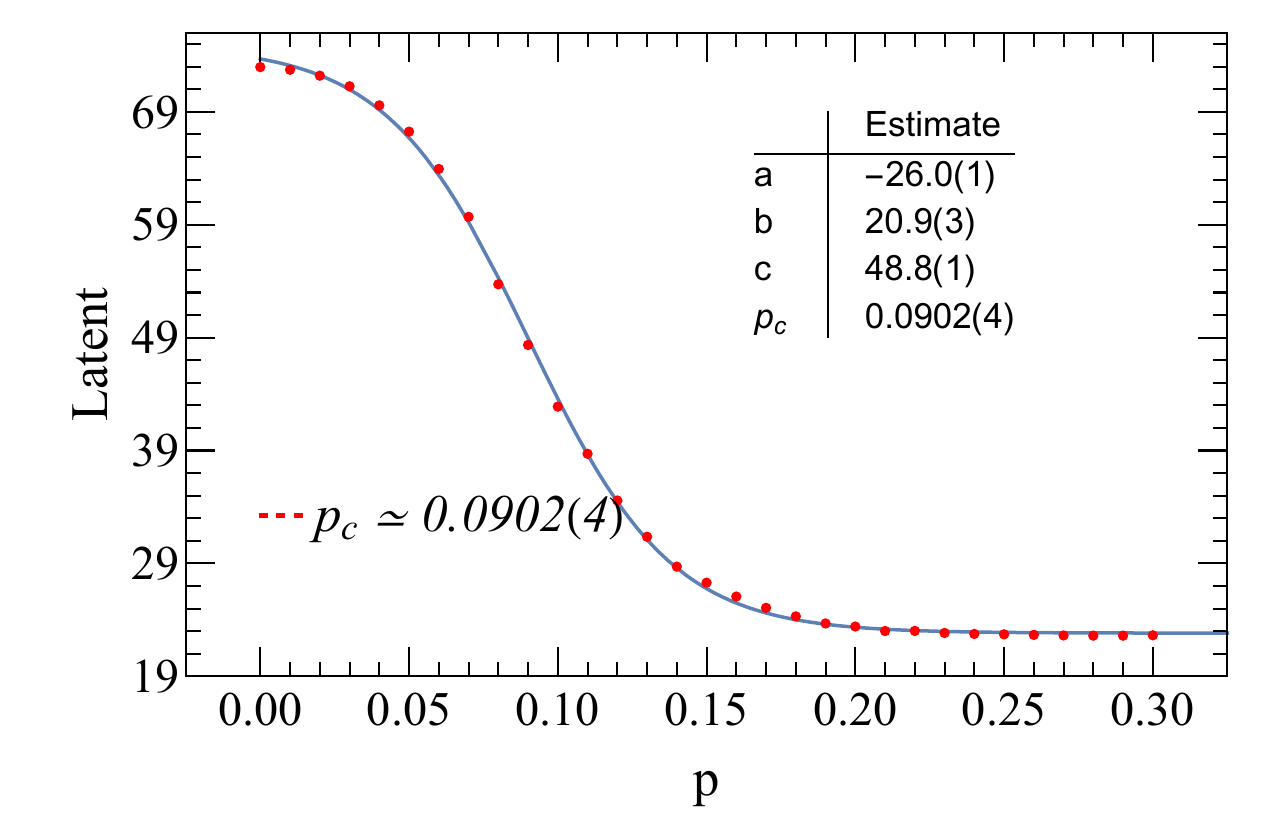}\\
    (a) &  (b) &  (c) \\
    \includegraphics[width=0.28\textwidth]{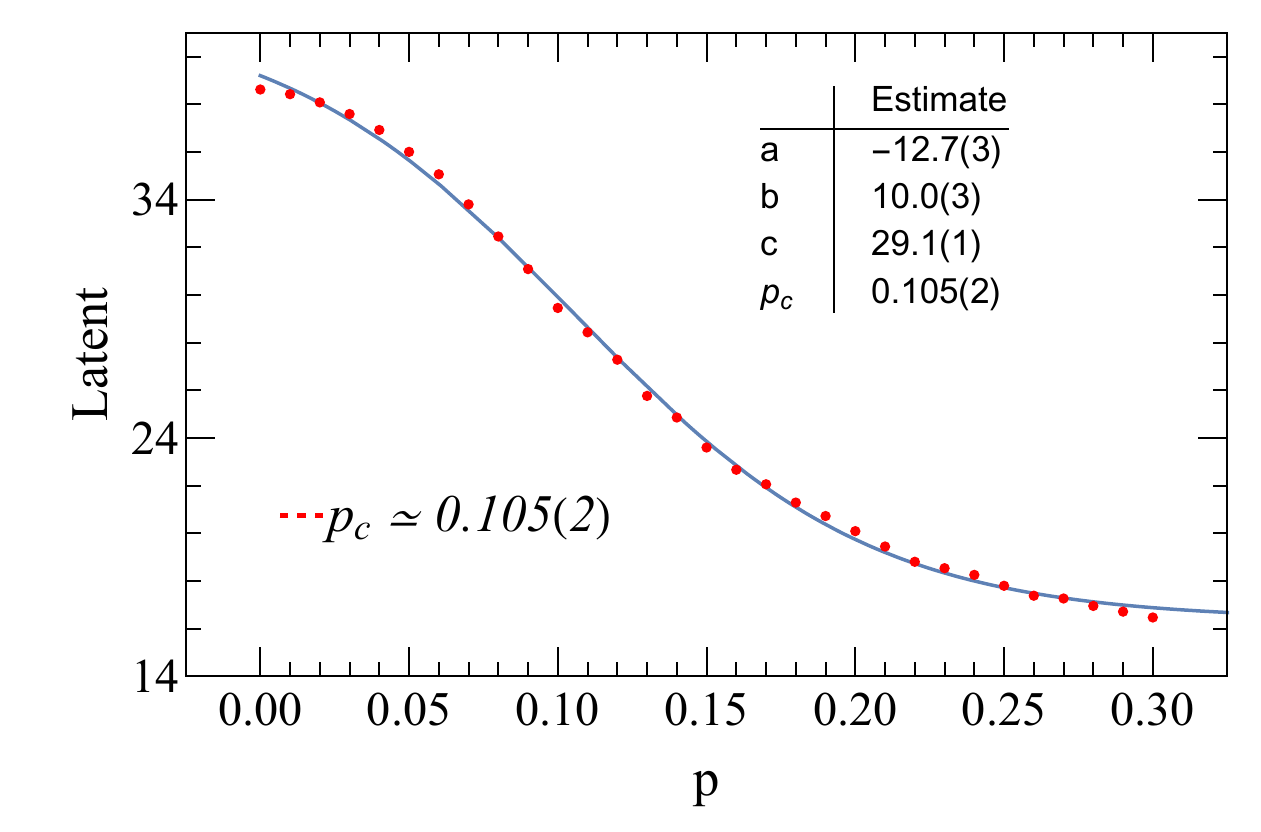} & \includegraphics[width=0.28\textwidth]{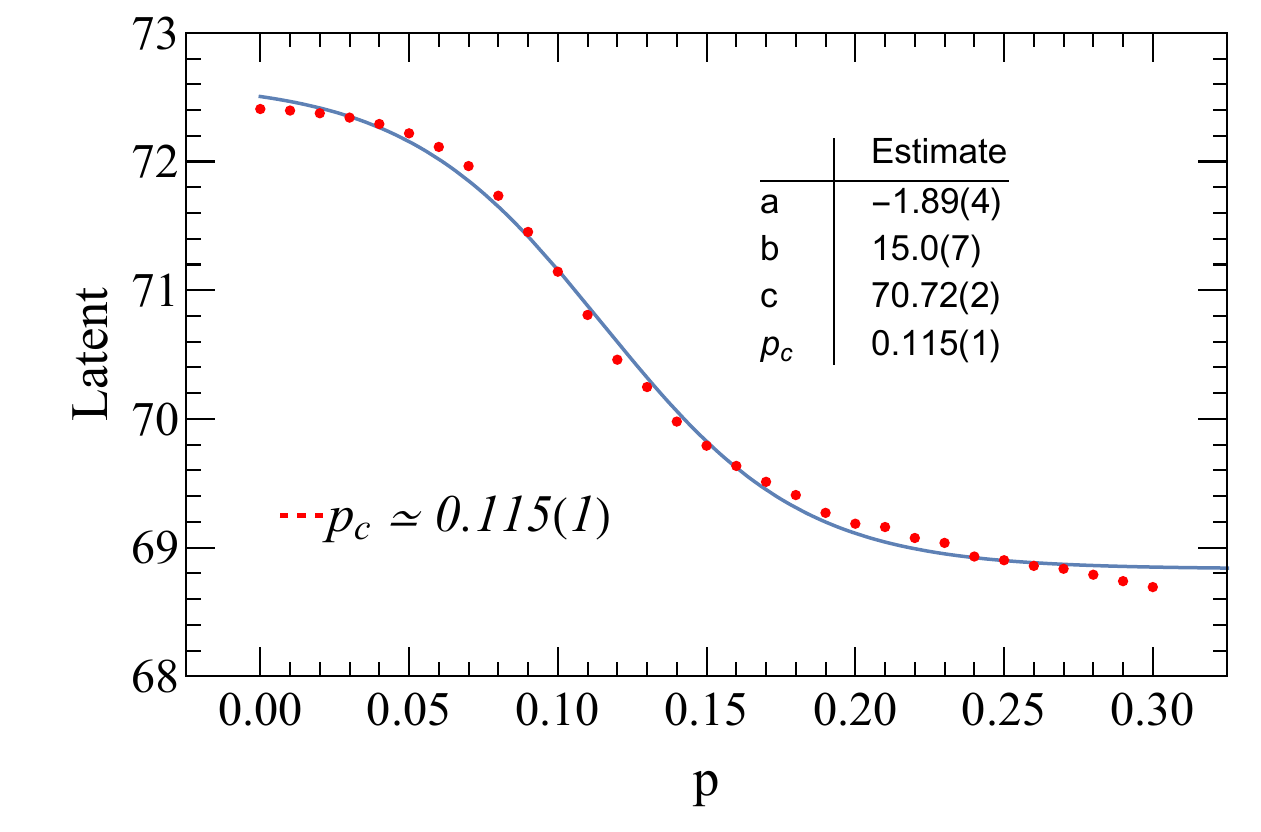} &
    \includegraphics[width=0.28\textwidth]{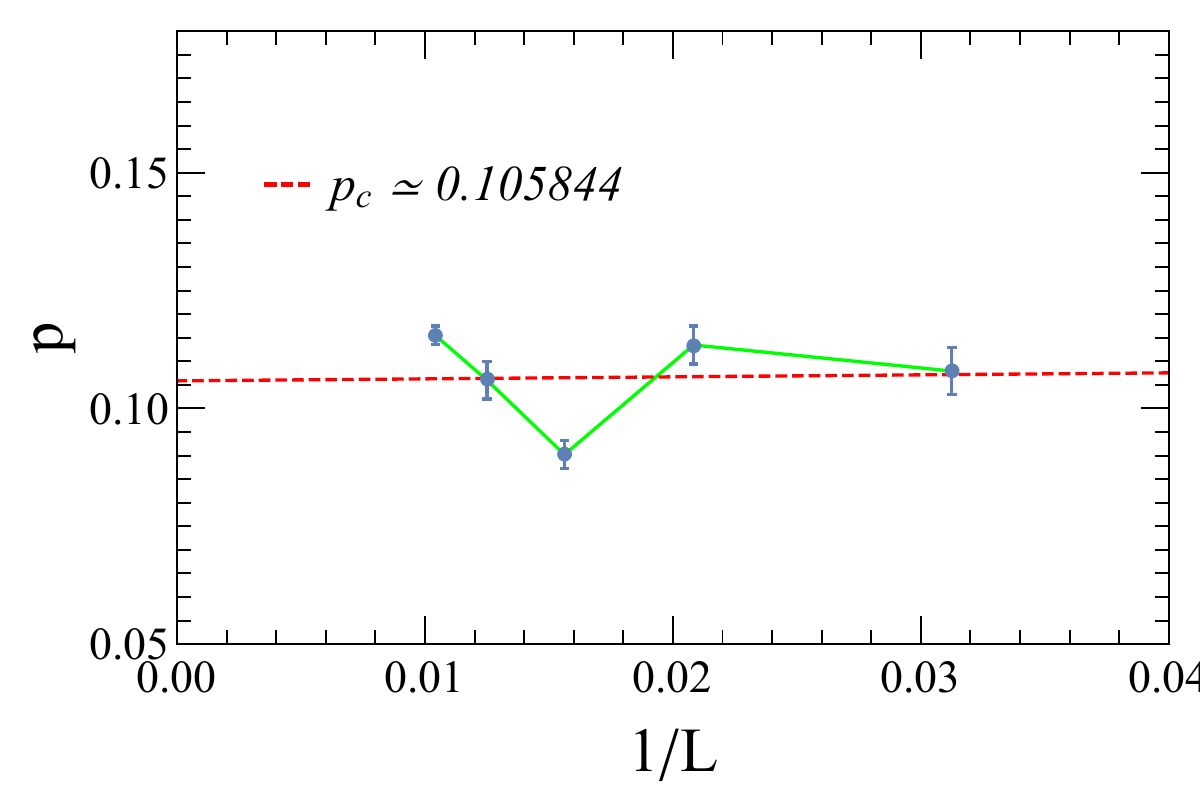} \\
    (d) & (e) & (f)
\end{tabular}
\caption{Autoencoder results of (1+1)-dimensional PCPD ($D=0.05$). \textbf{a-e}, encoding of the raw PCPD configurations, using a single hidden neuron activation $Latent$ which is a function of the annihilation probability. \textbf{f}, finite-size scaling, where $L=32,48,64,80,96$.}
\label{d005_pcpd_finite}
\end{figure*}

In the PCPD model, it includes not only the pairwise reaction between particles but also the diffusion motion of particles. First, we investigate the autoencoder learning results when $D = 0.05$. Fig.\ref{d005_pcpd_p1p2} (a) represents the scatter plot of hidden variables extracted from configurations of $10$ different $p$'s, and Fig.\ref{d005_pcpd_p1p2} (b) represents the counterpart of $41$ different $p$'s. For each of $p$, $100$ samples are generated. From these two panels, we can see that the configurations corresponding to the same $p$ are clustered in the proximity of one another. And all the clustered zones are well ordered according to the values of $p$. This allows for the conclusion that for (1+1)-dimensional PCPD, autoencoder can accurately cluster its configurations.

Furthermore, we hope to be able to obtain the threshold $p_{c}$ by limiting the number of potential neurons. Let the number of potential neurons in the autoencoder network be $1$, the plot of $Latent$ variable versus $p$ is shown in Fig. \ref{d005_pcpd_finite}. In order to achieve higher accuracy, a finite-size scaling is undertaken for the (1+1)-dimensional PCPD system at $D = 0.05$, where $L=32,48,64,80,96$. And the critical point is estimated to be 0.105844, quite close to $p_{c}=0.10439(1)$ as given in \cite{odor2003critical}.

As seen, the diffusion effect of $D = 0.05$ is not that significant. To study the cases when the diffusion rate is accelerated, we need to increase the value of $D$. Interestingly, the particle annihilation rate increases with the increase of $D$ and consequently, the number of occupied lattices drops rapidly. As $D$ increases up to a certain value, it seems that our autoencoder neural network cannot effectively detect the hidden structure of the system. We conjecture that larger $D$ could lead to a rapid decay of particle number in the system so that the neural network may not learn anything useful. That is to say, the input information to the neural network will be very limited, and the training is not effective without sufficient feeds. Our extensive studies  indicate that the capture of the PCPD critical point by autoencoder neural network is more feasible when $D \leq 0.5$ is satisfied.

\subsubsection*{Supervised learning of the PCP and the PCPD}
\label{Supervised_learning}
\begin{figure*}[htbp]
\begin{tabular}{cc}
   \includegraphics[width=0.4\textwidth]{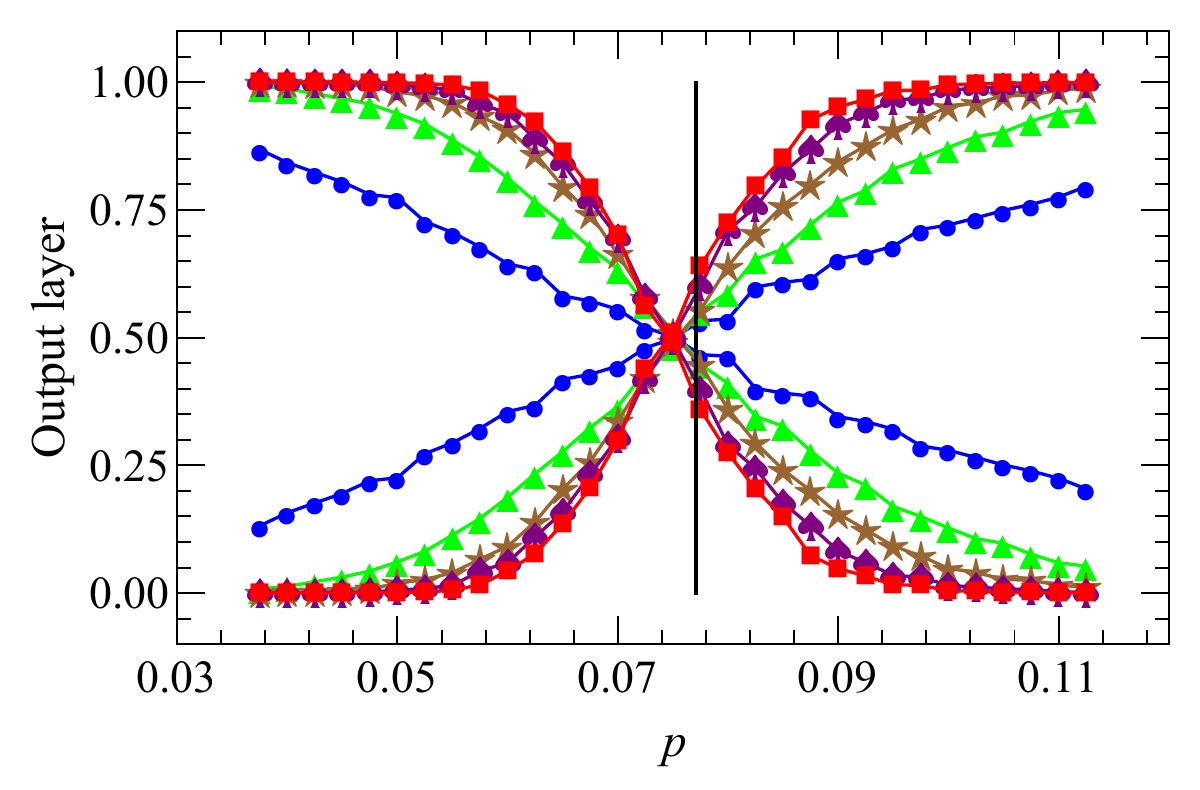} &
   \includegraphics[width=0.4\textwidth]{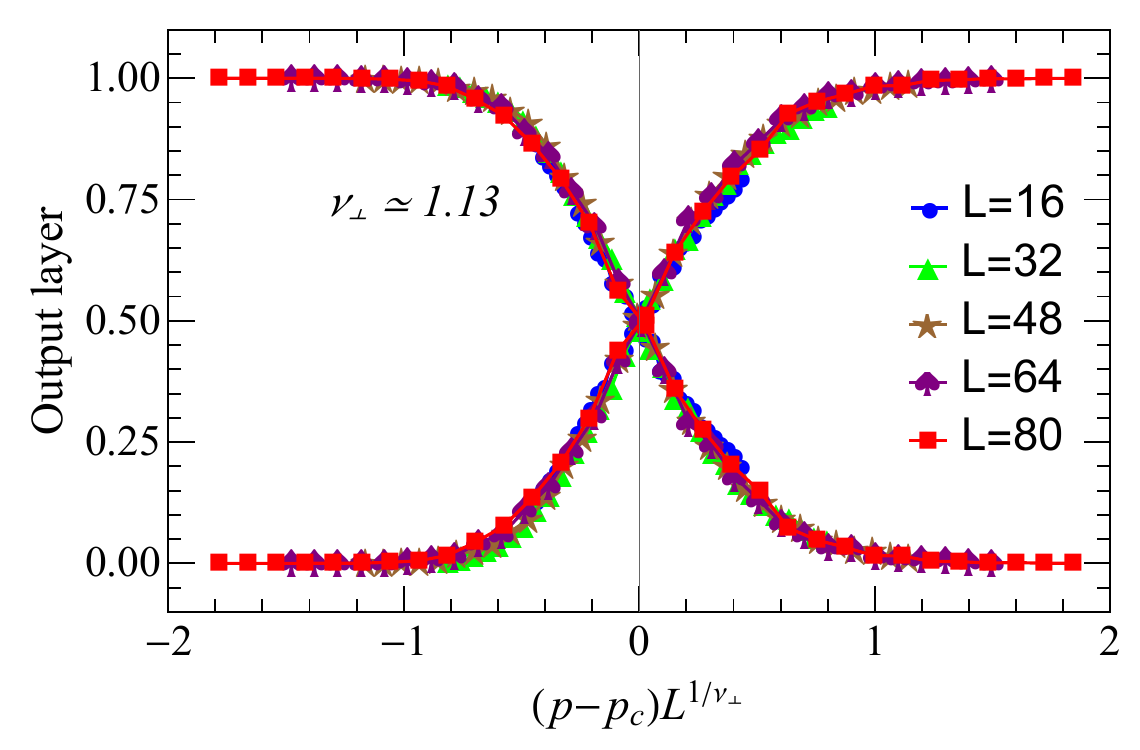} \\
    (a) & (b) \\
   \includegraphics[width=0.4\textwidth]{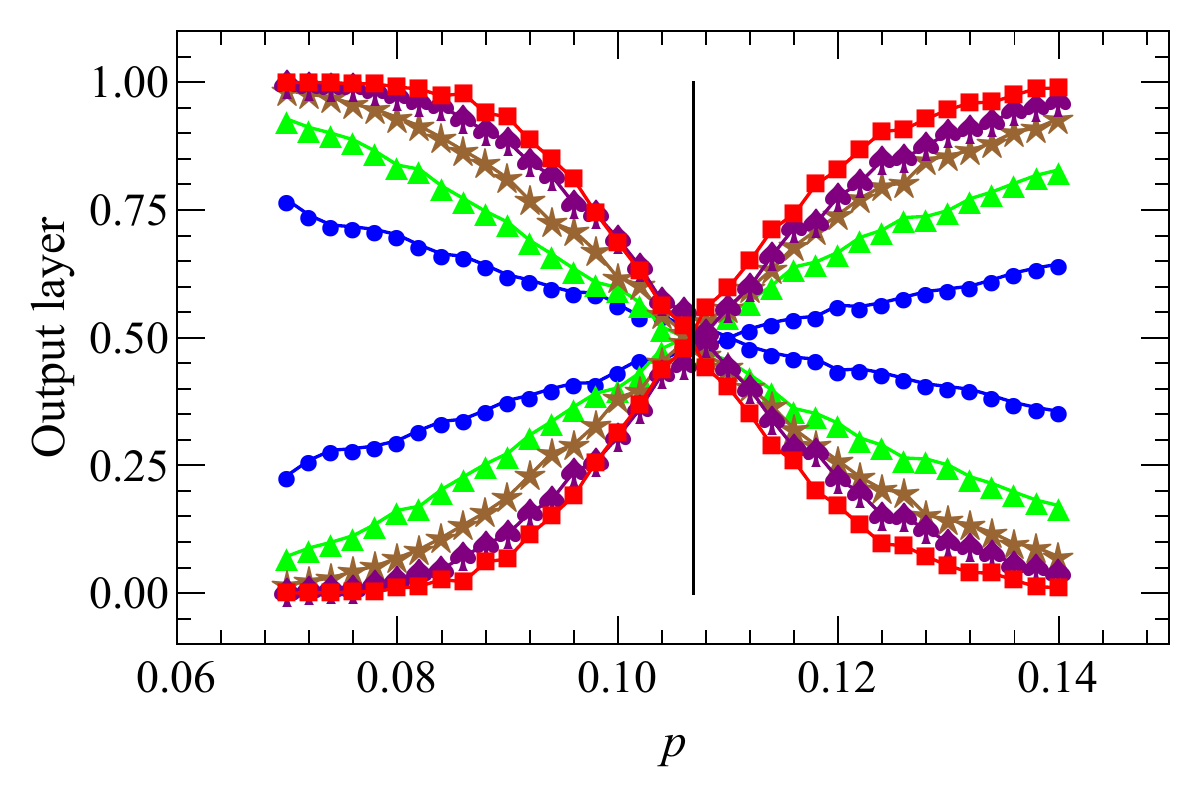} &
   \includegraphics[width=0.4\textwidth]{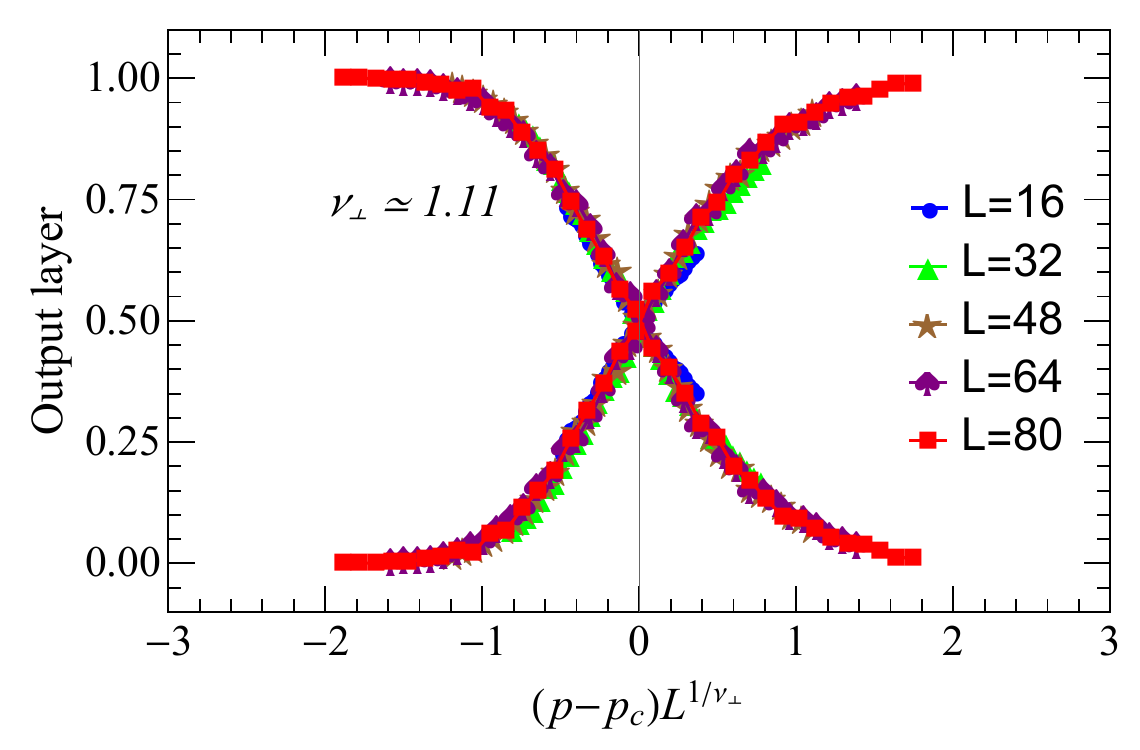} \\
    (c) & (d) \\
   \includegraphics[width=0.4\textwidth]{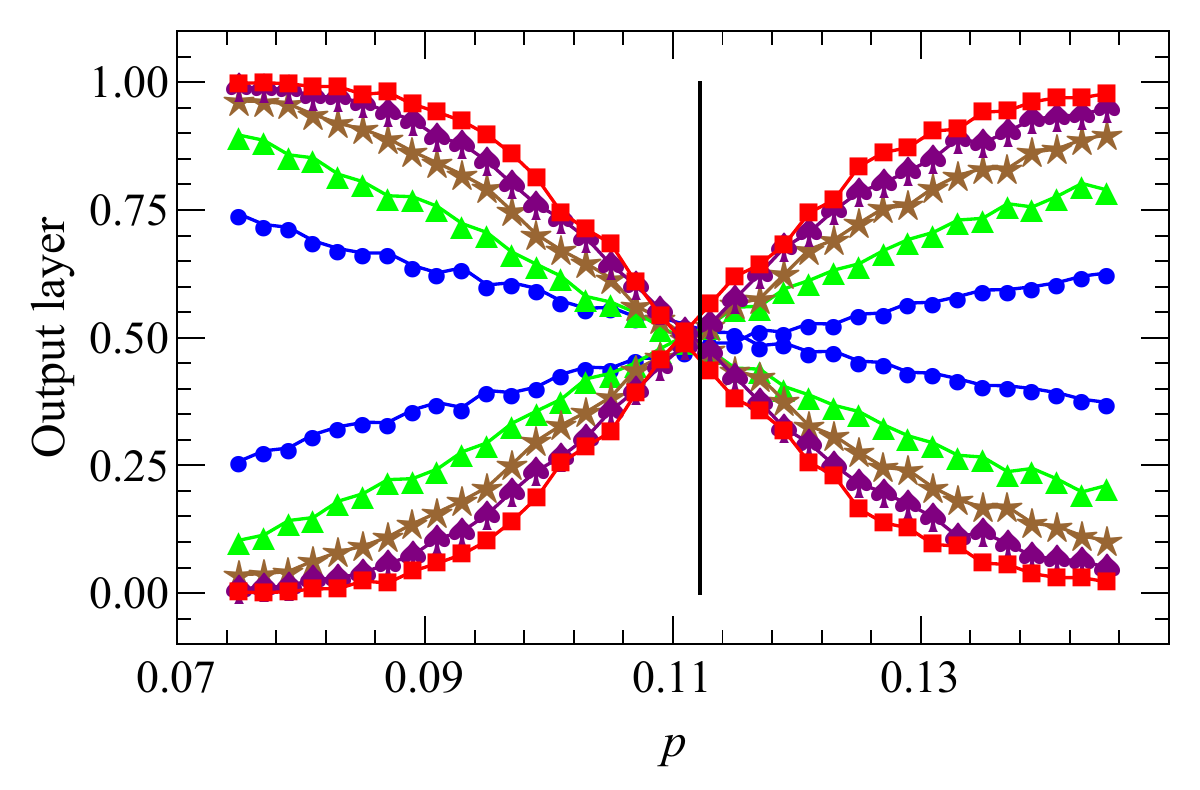} &
   \includegraphics[width=0.4\textwidth]{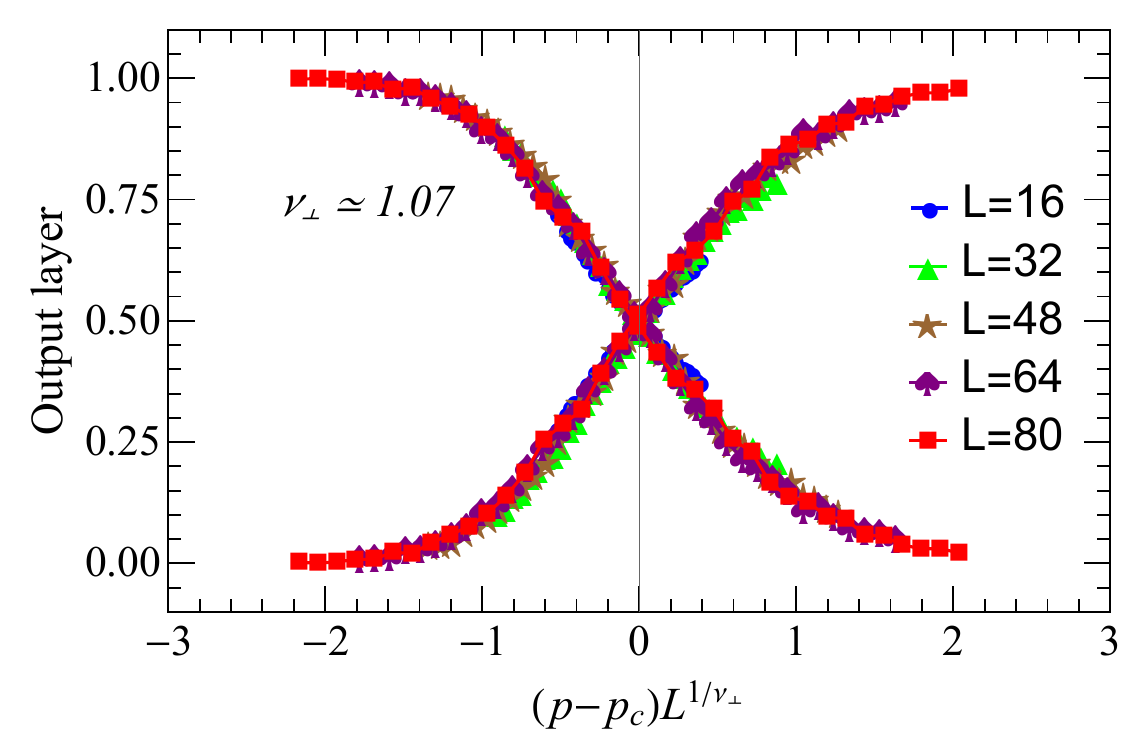} \\
    (e) & (f) \\
   \includegraphics[width=0.4\textwidth]{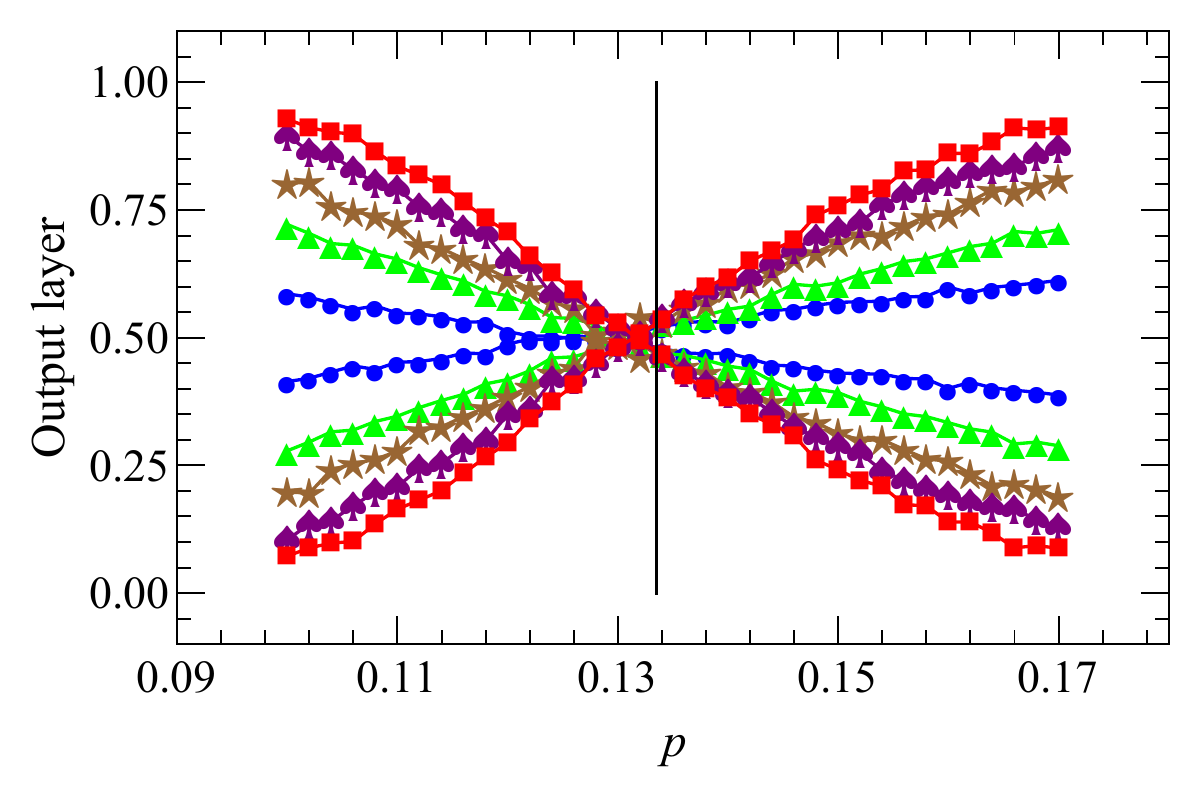} &
   \includegraphics[width=0.4\textwidth]{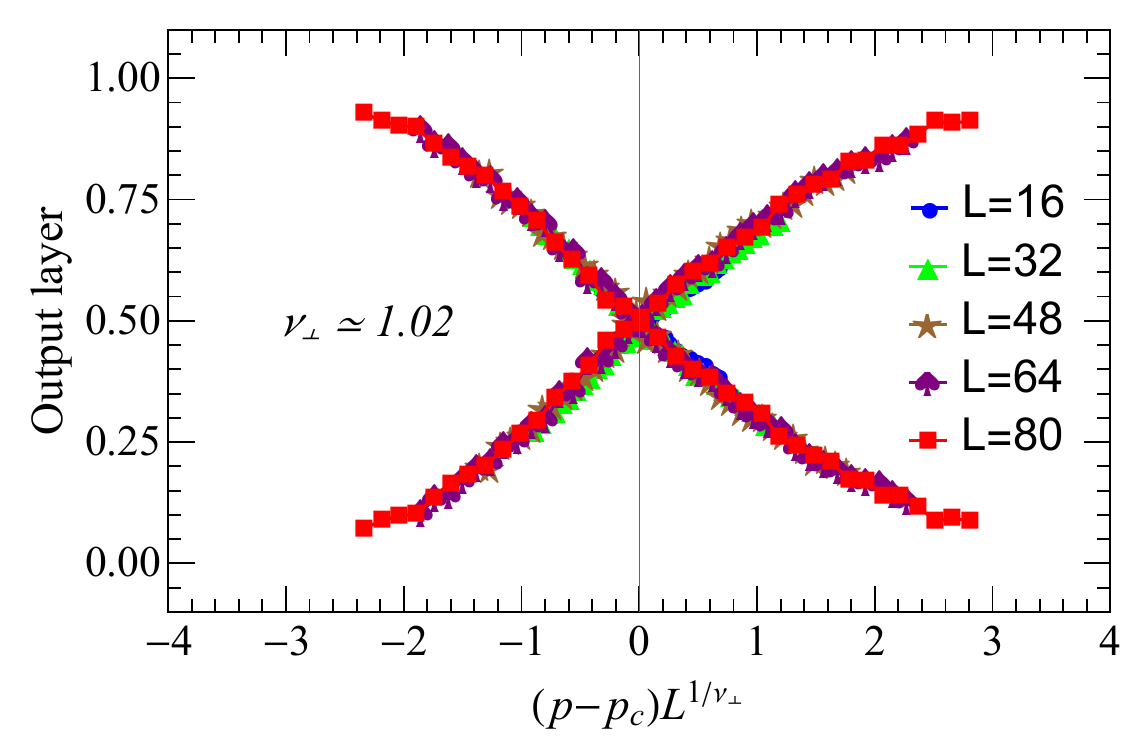} \\
    (g) & (h)
\end{tabular}
\caption{Supervised learning results of (1+1)-dimensional PCP and PCPD by FCN. \textbf{a}, The output layer, averaged over a test set, as a function of the bond probability $p$. \textbf{b}, Data collapse of the average output layer as a function of $( p- p_{c}(D)) L^{1/\nu_{\perp}}$. System sizes of $L =  16$, 32, 48, 64 and 80 are represented by different colors, respectively. \textbf{a} and \textbf{b} are the results of PCP, \textbf{c} \textbf{d}, \textbf{e} \textbf{f} and \textbf{g} \textbf{h} correspond to the results when the diffusion rate $D$ of PCPD is $0.1$, $0.2$, and $0.5$, respectively.}
\label{super_pcp_pcpd}
\end{figure*}

\begin{table*}[h]
	\centering
	\begin{tabular}{|c|c|c|c|c|c|c|}
        \hline
  D        & 0      & 0.05     &0.1    &0.2   &0.5 &0.7 \\
  \hline
 $\nu_{\perp}$(ML)  & 1.13(1) &1.12(1) & 1.11(1) &1.07(1) & 1.02(1) &0.99(1)  \\

        \hline
   $\nu_{\perp}$(MC($L=80$)) & 0.561(16) & 0.361(4) &  1.245(10) & 1.102(15) & 1.415(14) & 1.409(10) \\
        \hline
   $\nu_{\perp}$(MC \cite{odor2003critical,dickman2001pair}) & 1.092(7) & 0.775(10) &  0.830(7) & 0.873(8) & 1.009(16) & 1.02(2) \\
        \hline
   \end{tabular}
\caption{ML and MC simulation results of spatial correlation exponent $\nu_{\perp}$ with the diffusion rate $D$.}
\label{dvsniu}
\end{table*}

\begin{table*}[h]
	\centering
	\begin{tabular}{|p{2.5cm}<{\centering}|c|c|c|c|c|c|c|}
        \hline
 $\nu_{\perp}$  &0.91   & 0.96 & 1.01 &1.06 &1.09 &1.10 &1.11 \\
 \cline{1-1} \cline{2-2} \cline{3-3} \cline{4-4}  \cline{5-5} \cline{6-6} \cline{7-7} \cline{8-8}
 Euclidean Distance  & 0.1320    &0.1186   & 0.0982 &0.0746 & 0.0634 &0.0611 &0.0599 \\
  \cline{1-1} \cline{2-2} \cline{3-3} \cline{4-4}  \cline{5-5} \cline{6-6} \cline{7-7} \cline{8-8}
 $\nu_{\perp}$  &1.31  & 1.26 & 1.21 &1.16 &1.13 &1.12 &   \\
  \cline{1-1} \cline{2-2} \cline{3-3} \cline{4-4}  \cline{5-5} \cline{6-6} \cline{7-7} \cline{8-8}
 Euclidean Distance  & 0.1980  &0.1529   & 0.1090 &0.0732 &0.0615 &0.0601 & \\
        \hline
   \end{tabular}
\caption{The Euclidean distance between two sigmoid curves varies with the spatial correlation exponent $\nu_{\perp}$. }
\label{niu_eucliddistance}
\end{table*}

\begin{figure*}[h]
\begin{minipage}{0.3\linewidth}
  \centerline{\includegraphics[width=5.0cm]{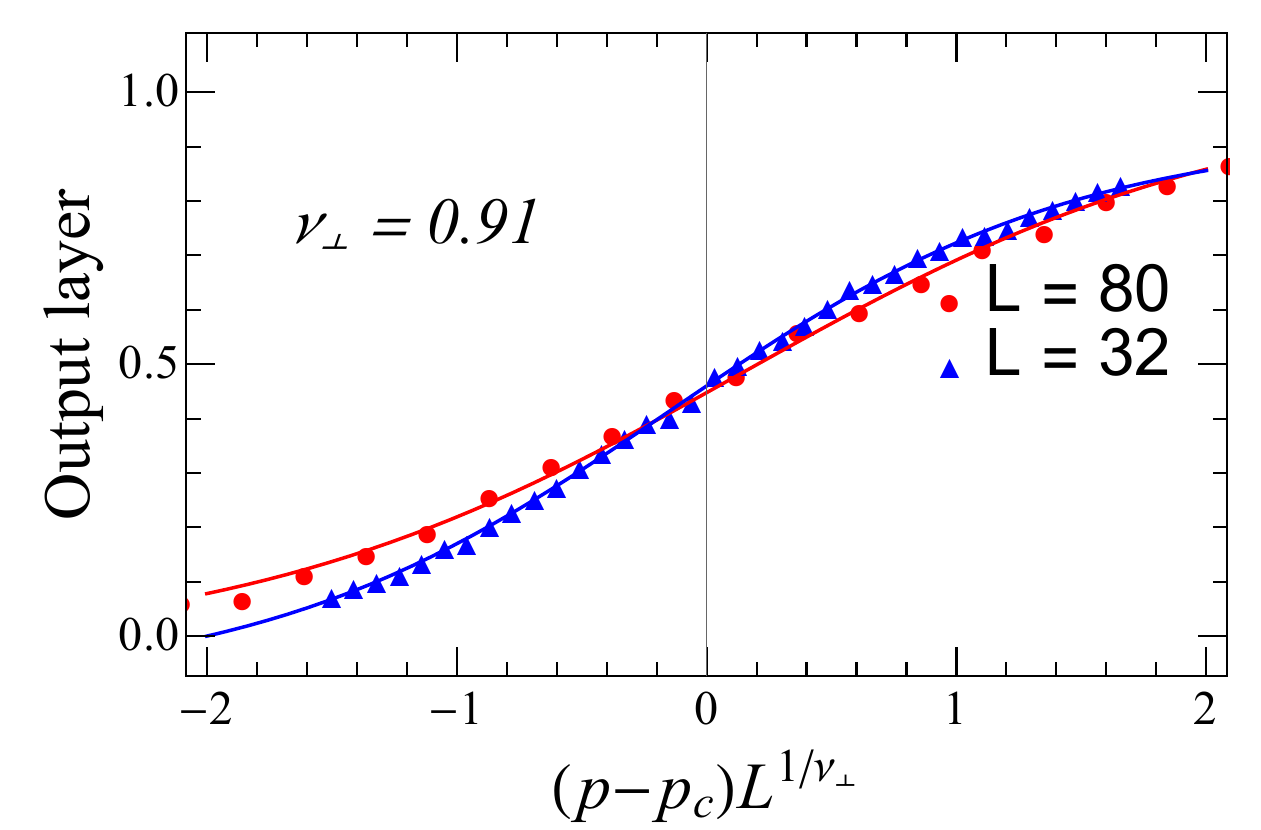}}
  \centerline{a}
\end{minipage}
\hfill
\begin{minipage}{0.3\linewidth}
  \centerline{\includegraphics[width=5.0cm]{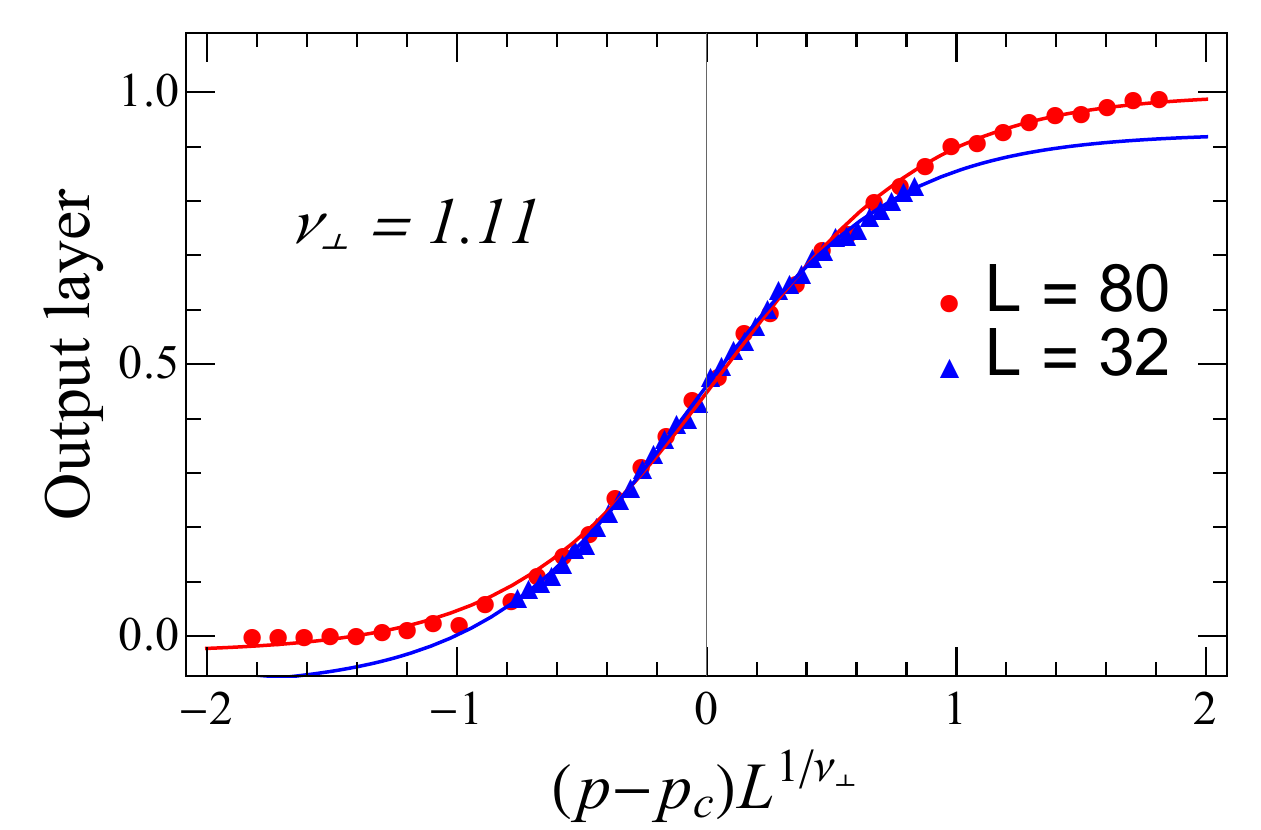}}
  \centerline{b}
\end{minipage}
\hfill
\begin{minipage}{0.3\linewidth}
  \centerline{\includegraphics[width=5.0cm]{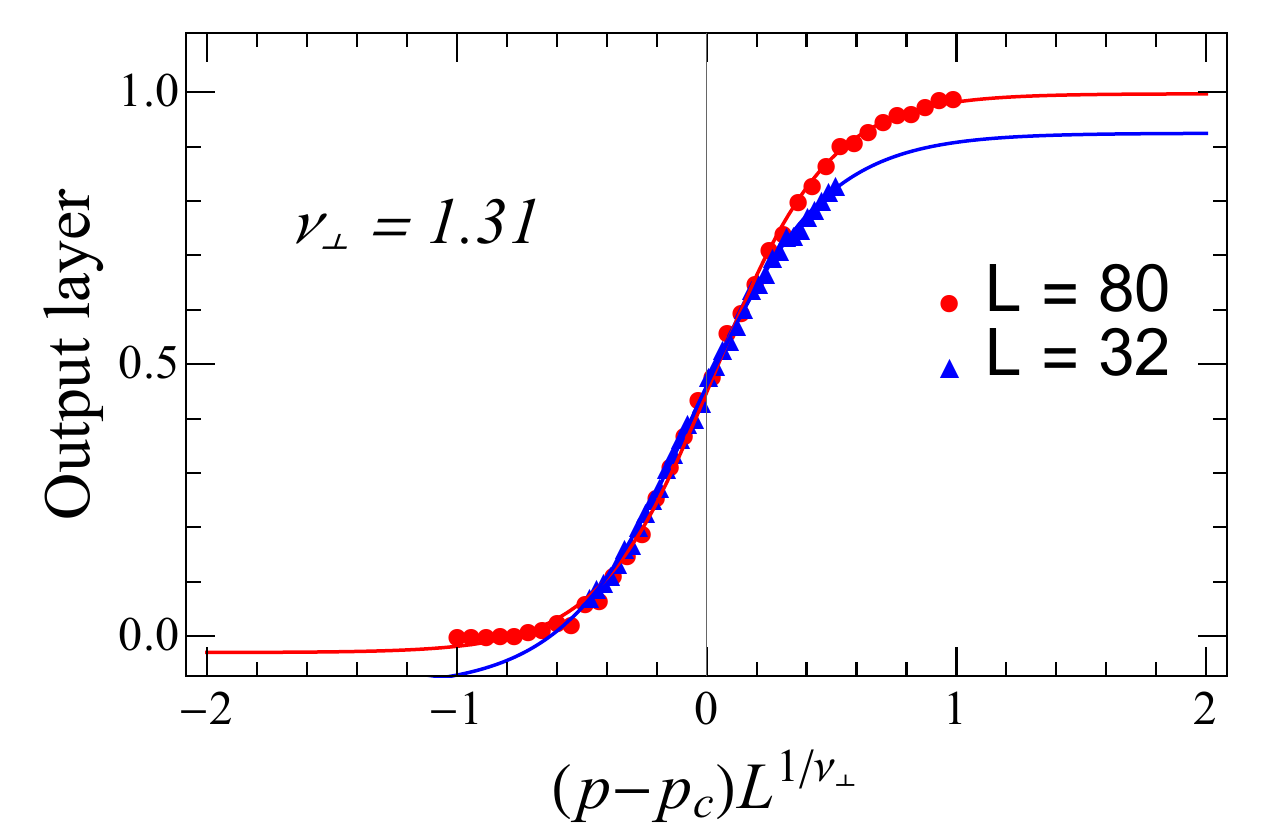}}
  \centerline{c}
\end{minipage}
\caption{\textbf{a} and \textbf{c} are results of data collapse with different $\nu_{\perp}$ in the PCPD, where diffusion rate is $D = 0.1$.}
\label{3niu_datacollapse}
\end{figure*}

Unsupervised learning can detect the transition points of critical systems, whereas supervised learning can yield some critical exponents by rescaling or the so-called data collapse. There are two correlation exponents in non-equilibrium phase transitions, called spatial and temporal correlation exponents, respectively. Here we focus primarily on the former, spatial correlation exponent. Through the dynamical exponent $z = \nu_{\parallel} / \nu_{\perp}$, the temporal correlation exponent can be indirectly measured.

Here in supervised learning, we employ the fully connected network (FCN) to identify phases of the PCPD. The basic architecture of FCN used in this paper can be found in one of our previous articles \cite{shen2021supervised}. The hyper-parameters we use are as follows: the learning rate is $0.0001$, the batch size is $1024$ and the regularization parameter is $0.01$. The mean cross-entropy between the label vector $y_{l}$ and the output layer vector $y$ used in our FCN also contains a L2-norm mentioned above to prevent over-fitting. Our FCN contains $100$ neurons in the hidden layer, and two in the output layer. Input configurations are labelled by "0" for probabilities of annihilation less than the critical threshold, and "1" for rest probabilities. After the neural network is trained and fine tuned, we obtain the learning results of the PCP and the PCPD, shown in Fig. \ref{super_pcp_pcpd}. The first column of Fig. \ref{super_pcp_pcpd} represents the output results of two neurons of five different system sizes, and the intersections are the transition points predicted by the learning. Obviously, these predictions are consistent with the theoretical thresholds, and the details of the learning procedure can refer to \cite{shen2021supervised}.

The second column of Fig. \ref{super_pcp_pcpd} displays the data collapses in which the horizontal co-ordinates have been rescaled by a factor related to the spatial correlation exponent $\nu_{\perp}$. Both PCP and PCPD are binary reaction diffusion processes, and the PCP belongs to the universality class of the DP, whereas the university class of PCPD remains vague. The spatial correlation exponent given by Fig. \ref{super_pcp_pcpd} (b) is $\nu_{\perp} \simeq 1.13(1)$, which approaches the theoretical one $\nu_{\perp} \simeq 1.09 $. For the PCPD its critical threshold $p_{c}(D)$ may depend on the diffusion rate. $\nu_{\perp}$ is also found to be dependent on $D$, the diffusion rate, as give in Table \ref{dvsniu}. Especially $\nu_{\perp}$ decreases as $D$ increases. Compared with the results obtained by Monte Carlo simulations in Refs. \cite{odor2000critical,dickman2002nonuniversality,odor2003critical}, the measurements of $\nu_{\perp}$ are accompanied by certain amount of uncertainties, caused by various factors.

Although our measurements show that $\nu_{\perp}$ may vary with the diffusion rate, we have to be very cautious in drawing such a conclusion, due to the fact that ML so far only works with small systems. Because there are finite-size effects, as well as fluctuations driven by diffusion. Therefore, the dependence of $\nu_{\perp}$ on $D$ might be an artifact caused by the combination of finite-size effects and diffusion-driven fluctuations.

Generally speaking, ML of phase transitions has been very well adapted to small-sized systems. For comparison, MC simulations are also performed for (1+1)-dimensional PCPD of same sizes that have been set for ML. The critical exponents $\nu_{\perp}$ for different diffusion rates are given in Tab. \ref{dvsniu}, obtained from ML, MC small systems and MC literatures, respectively. According to the method in \cite{odor2003critical}, we need to know $\beta$ and $\beta / \nu_{\perp}$ to get $\nu_{\perp}$. Dealing with the effective $\beta$ exponent, we refer to formula
\begin{equation}
\beta_{eff} = \dfrac{\ln(\rho(\infty,\epsilon_{i})) - \ln(\rho(\infty,\epsilon_{i-1}))}{\ln(\epsilon_{i}) - \ln(\epsilon_{i-1})},
\label{beta_eff}
\end{equation}
where $\rho(\infty,\epsilon_{i})$ means steady state densities, $\epsilon_{i} = p_{c} - p_{i}$. We choose five different diffusion rates to calculate effective $\beta$ exponents. In this part of MC simulations, the model parameters selected are as follows. The lattice size of the PCPD system is $L=80$, the total time step is $t=6400$, and the number of ensemble average is $1000000$ for each annihilation probability. And we find that almost all the extrapolation values of $\beta_{eff}$ are larger than those in \cite{odor2003critical}. This is most likely due to strong finite-size effects or the uncertainties of p-values used. In calculating $\beta$, we use logarithmic correction methods like in \cite{odor2003critical}, defined as
\begin{equation}
\rho(\infty,\epsilon) = [\epsilon/(a+b \ln(\epsilon))]^{\beta}.
\end{equation}
For calculating $\beta / \nu_{\perp}$, we refer to the quasi-steady state density (averaged over surviving samples)
\begin{equation}
\rho_{s}(\infty,p_{c},L) \propto L^{-\beta/\nu_{\perp}},
\label{rou_s}
\end{equation}
where the lattice size is $L=16,32,48,64,80$, the total time step is $t=L^{z'}$(we choose $z'=2$), and the number of ensemble average is $1000000$ for each annihilation probability. Since the critical annihilation probability $p_{c}$ of the small-sized PCPD system fluctuates greatly, we obtain a critical value $p_{c}=0.102(3)$ with diffusion rate $D=0.1$ in the $L=80$ system. In our simulations, something interesting happens. If we choose an accurate critical value $p_{c} = 0.10688$ \cite{odor2003critical} obtained from large size ($L=10^{5}$), $\beta_{eff}$ is inaccurate but $\beta / \nu_{\perp}$ is relatively reasonable. On the contrary, if we choose the effective critical value $p_{c} = 0.102$ estimated from small size ($L=80$), $\beta / \nu_{\perp}$ is not correct but $\beta_{eff}$ is relatively reasonable. In short, the choice of both $p_{c}$ values will result in a large deviation for calculating $\nu_{\perp}$ in a small-sized system. In our work, $p_{c} = 0.10688$ \cite{odor2003critical} is taken in calculating $\nu_{\perp}$.

Unsurprisingly, the MC simulations of small-sized systems showed larger deviations compared to the counterparts of large-sized systems. These deviations are caused partly by the randomness of diffusion, and partly by the finite sizes. But supervised ML results are revealed to be relatively stable and much less affected by the finite sizes. We believe that supervised ML to calculate the critical exponents can provide some reference for the study of the PCPD. Of course, our algorithm and computation scale can still be improved, which will be the work of the future.

Due to its strong correction to scaling, the PCPD is notoriously known for its extremely slow crossover behavior to the scaling region, rendering the estimations of both its critical point and the critical exponents hard (see e.g.~the extensive review \cite{henkel2004non}). Therefore, even though several studies, such as a bosonic variant of the PCPD \cite{kockelkoren2003absorbing} and the refined mean-field phase portrait analysis \cite{elgart2006classification}, suggested that the PCPD constitutes a novel universality class different from the DP or the PCP, most Monte Carlo simulation studies have yet defied this conclusion. Most notably, many elaborate simulations reveal that its critical properties seem to depend on the considered diffusion rate \cite{odor2000critical,dickman2002nonuniversality,de2006moment}. Nevertheless, by sophisticatedly taking into account the effects of correction to scaling, a more recent study \cite{park2014critical} was able to obtain a diffusion-independent decay exponent that is markedly distinct from the DP value. Given the rather limit system size and simulation time we used in the ML scheme, such large-scale, long-time behaviors as in Ref. \cite{park2014critical} of course can not be observed, but as in Refs. \cite{odor2000critical,dickman2002nonuniversality,de2006moment}, one should expect to observe a diffusion-dependent measurement for the spatial correlation exponent $\nu_{\perp}$ as opposed to other phase transition models.


Very little can be traced in the literature on how to estimate critical exponents from data collapse. Mostly the procedure relies on the eyes, without any solid foundation or reliable criterion. Here, we propose a more reliable means to determine $\nu_{\perp}$, combining the Euclidean distance. Take as an example the case where the diffusion rate is 0.1. We choose the two results of the output layers corresponding to $L = 32$ and $L = 80$, respectively. For binary classification neural networks, usually, the outputs obey sigmoid functions. After rescaling the abscissa, we perform sigmoid function fitting for the two curves. Assume $Y = F_{1}(x)$ and $Z = F_{2}(x)$, then the Euclidean distance is $Euclidean Distance = \sqrt{\sum \limits_{i = 1}^{n} (Y_{i} - Z_{i})^{2}}$. When calculating the Euclidean distance, we uniformly select all $X$'s in the range of $[-0.5,0.5]$ with an interval of $0.02$.  The Euclidean distances are provided in Table \ref{niu_eucliddistance}, and the data collapse results for three different $\nu_{\perp}$'s are plotted in Fig. \ref{3niu_datacollapse}. Starting with $\nu_{\perp} = 0.91$, with the increase of $\nu_{\perp}$, the Euclidean distance first decreases until $\nu_{\perp} = 1.11$, and then increases after that. $\nu_{\perp} = 1.11$ is the value that we are looking for, which corresponds to the optimal fitting. This method successfully avoids the instability of naked eyes, and therefore might be served as a methodology for data collapse.

\subsubsection*{Discussion}
Since the parameters built in the autoencoder are obtained by minimizing the cost function at certain training samples, it's necessary to check its prediction ability. Here, we implement the cross-entropy as the cost function, and compare its values at the last training epoch in training set to those in validation set. The values of different system sizes are displayed in Table \ref{train_validation_accuracy}. It can be found that for most sizes the two losses, namely train loss and validation loss, are quite close to each other, and the peak of their difference, around $5\%$, appears at $L=32$. Therefore, the trained network is believed to be applicable for all configurations.

Now that the training loss reveals reconstruction ability of the autoencoder, it's still interesting to discuss the relation between training loss and lattice size. Firstly, the training loss raises its value from $L=16$ to $L=48$. This is because the relation between sites of configurations gets augmented as the lattice size increases, but it's insufficient for autoencoder to generate an appropriate encoder-decoder chain. Once $L$ is larger than 64, the loss begins to fall, which implies that a larger lattice size is needed to reduce training loss. This upward and downward trend should be able to demonstrate the finite-size effect of the system. 

\begin{table*}[h]
	\centering
	\begin{tabular}{|c|c|c|c|c|c|c|}
	        \hline
       lattice size L    & 16  & 32  &  48   &  64  & 80 \\
        \hline
training loss    & 0.5564  & 0.6564  &0.6636   &0.6397 & 0.6182 \\
        \hline
validation loss  & 0.5404  & 0.6237  &0.6568   &0.6076 & 0.6240 \\

        \hline
   \end{tabular}
\caption{The training and validation loss of autoencoder for (1+1)-dimensional PCPD with the diffusion rate $0.1$, where each annihilation probability in training and validation sets corresponds to a sample number of $2000$ and $200$.}
\label{train_validation_accuracy}
\end{table*}

Note that the current ML is only feasible for systems of small sizes, which certainly brings up finite-size effects. Additionally, PCPD itself is a very special non-equilibrium phase transition model, in which there is diffusion of particles which is also affected by the system size. From the perspective of ML, in order to reduce errors to a greater extent, what we can do is to optimize various parameters of the neural network, expand the training scale and increase the number of test samples. We hope this could help improve the computational accuracy.

\begin{figure}[h]
\centering
\includegraphics[width=0.6\textwidth]{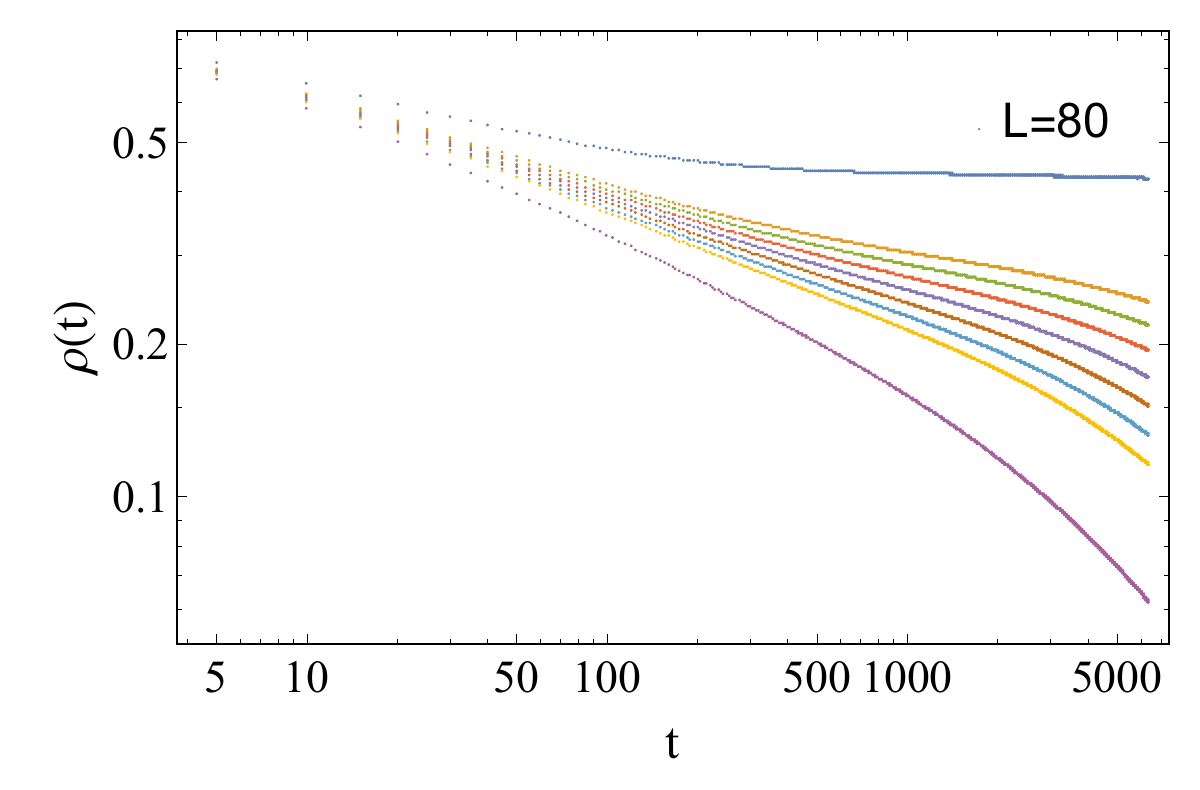}
\caption{(1+1)-dimensional PCPD Monte Carlo simulations, where the lattice size is $L=80$, the total time step is $t=6400$, and the number of ensemble average is $1000000$, for different annihilation probabilities of $p=0.092,0.100,0.101,0.102,0.103,0.104,0.105,0.106,0.111$, respectively.}
\label{pc_80_01}
\end{figure}

We hope to find an available benchmark or a baseline level for the desired predictive accuracy. Here, we calculate the critical value of small-sized PCPD by MC simulations, and the MC results of critical values are shown in Tab. \ref{ml_mc_pcpd_pc}. It can be seen from Fig. \ref{pc_80_01} that in the (1+1)-dimensional PCPD with lattice size $L=80$, it seems difficult to find a clear power-law density decay curve near the critical regime. Then we find that the predictions of small-size systems of ML are slightly less accurate than those of large-size systems of MC \cite{odor2003critical}, although the deviation is not that large. Therefore, we believe that ML can provide some reference for critical value prediction of the PCPD model.

\begin{table*}[h]
	\centering
	\begin{tabular}{|c|c|c|c|c|c|c|}
        \hline
  lattice size        & 32    & 48     &64   &80    &96 \\
        \hline
$p_{c}$(MC)     & 0.090(10) & 0.095(5) & 0.099(4) & 0.102(3) & 0.102(2)   \\
        \hline
   \end{tabular}
\caption{For (1+1)-dimensional PCPD, the comparison of critical point $p_{c}$ on the diffusion rate $D = 0.1$ between ML and MC simulations with different lattice sizes. The theoretical value is $p_{c} = 0.10439(1)$ \cite{odor2003critical}.}
\label{ml_mc_pcpd_pc}
\end{table*}

\section*{Conclusion}
In this paper, we apply unsupervised learning methods (PCA and autoencoder) and supervised learning to the binary process PCPD of non-equilibrium phase transition models. The main conclusions are as follows.

First, using PCA, the linear dimensionality reduction method, we can cluster different configurations and visually reduce the dimension of the PCPD. In addition, the first principal component after PCA dimensionality reduction can represent the order parameter of the model, that is, particle density. Via $p_{1}$ versus $p$, we can estimate the threshold of the model, which is consistent with the theoretical value.

Second, by using the convolutional autoencoder neural network, we can extract the feature information of original configurations of the PCPD through the compressed representation of hidden neurons. When the number of hidden neurons is $2$, we can have a prominent clustering representation. When the number is limited to $1$, the neuron can be treated as the order parameter of the model, that is, the particle density.

Third, the spatial correlation exponents are obtained by supervised learning of pair-contact process with different diffusion rates. The results alone, however, are not sufficient to indicate that the critical exponents $\nu_{\perp}$ of the PCPD are dependent on the diffusion rate $D$. Given the limitations of ML, there is no enough evidence that the PCPD conveys a new kind of absorbing phase transition. In addition, we propose a numerical method to obtain the correlation exponent with higher accuracy, by calculating the Euclidean distance between two fittings.

The present findings confirm that for the random reaction process of non-equilibrium lattice models, even if the model contains diffusion motion, as long as the evolution process of the particle configurations has a trend of change, PCA and the autoencoder neural network can extract or capture this feature so that we can quantify it and describe it.

So far, we have known that ML is applicable to the DP and PCPD models. There are still many unanswered models in the field of non-equilibrium lattice systems. ML can provide available help in understanding these models. In future investigations, ML techniques might be possible to reveal more information in studying statistical physics problems.

\section*{Appendix A: Principal component analysis}
For convenience, let's assume that the original data matrix $X_{m \times n}$ is decentralized. We want to get the reduced dimension representation, or the principal component representation, by the linear transformation $Y=XW$. $W=(w_{1},w_{2},..., w_ {n})$ is a new set of uncorrelated basis vectors we are looking for, and $w_{\ell}$ is the weight vector of the principal component. We can write it in the following form
\begin{equation}
X^{T}Xw_{\ell} = \lambda_{\ell}w_{\ell}.
\end{equation}
The eigenvalues corresponding to the weight vector $w_{\ell}$ are sorted in descending order, with $\lambda_{\ell} \geq \lambda_{2} \geq \lambda_{N} \geq 0$. The normalized eigenvalue $\widetilde{\lambda}_{\ell} = \lambda_{\ell}/\sum_{\ell=1}^{N}\lambda_{\ell}$ represents the explained variance ratio. In most cases, the first few principal components can represent most of the information of the original data.

In the SVD method, $X_{m \times n}$ can be any matrix, which we decompose as follows
\begin{equation}
X = U\Sigma V^{T},
\label{svd_equation}
\end{equation}
where $U$ is an $m*m$ square matrix of left singular vectors, $\Sigma$ is an $m*n$ singular value matrix, $V^{T}$ is an $n*n$ transposed matrix of right singular vectors. We express the dimensional change of the matrix as
\begin{equation}
X_{m*n} = U_{m*m}\Sigma_{m*n} V^{T}_{n*n}.
\end{equation}

According to an formula
\begin{equation}
X^{T}Xv_{\ell} = \lambda_{\ell}v_{\ell},
\end{equation}
where $v_{\ell}$ is the right singular vector. Combining this equation and Eq. \ref{svd_equation}, it is not difficult to prove
\begin{equation}
\sigma_{\ell}=\sqrt{\lambda_{\ell}}.
\end{equation}
In most cases, the sum of the singular values of the top $10\%$ or even $1\%$ accounts for more than $99\%$ of the total singular values. The right singular matrix can be utilized for column compression
\begin{equation}
X^{'}_{m*k} = X_{m*n} V^{T}_{n*k}.
\end{equation}
Comparing this expression with the PCA decomposition of $Y=XW$, one finds that the orthogonal matrix $V$ in SVD is exactly the orthogonal matrix $W$ in PCA. Therefore, the method based on the eigenvalue decomposition covariance matrix is a particular case of the SVD method. That is, the original matrix is square.

The code implementation of PCA is easy. The kernel function used by the Scikit-learn package in Python is SVD, which we can call with little hindrance.

\section*{Appendix B: Autoencoder}
\begin{figure}[ht]
\centering
\includegraphics[width=0.6\textwidth]{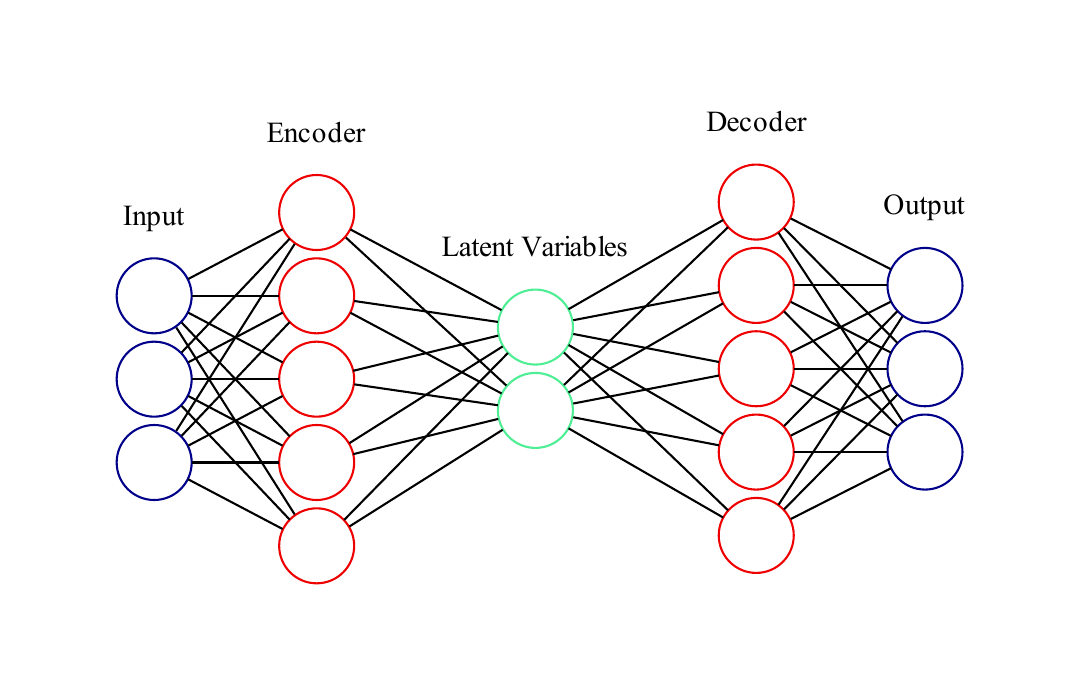}
\caption{ Neural network schematic structure of autoencoder.}
\label{autostruc}
\end{figure}

The encoder process of the autoencoder is considered as the non-linear enhanced version of PCA. An autoencoder (in Fig. \ref{autostruc}) is a neural network whose target output is its input without labeling the input samples. The learning goal of the autoencoder is to minimize reconstruction errors. In other words, it learns an approximate identity function such that the output $\hat{x}$ is approximately equal to the input $x$. As the network representation form of an autoencoder, we can use it as a layer to build a deep learning network. With appropriate dimensions and sparse constraints, the autoencoder can perform better than PCA and other technologies.

Autoencoders are data-specific, which means they can only compress data similar to what they have been trained to do. For example, an autoencoder trained on images of elephants would do poorly at compressing images of flowers, because the features it would learn would be specific to the elephant.

To build an autoencoder, we need three things that are an encoder function, a decoder function, and a loss function. The loss function is the amount of information lost between the compressed and decompressed data representations. Where the encoder process from the input layer to the hidden layer is as follows
\begin{equation}
h = g_{\Theta_{1}} (x) = \sigma(W_{1}x + b_{1})
\end{equation}
The decoder process from hidden layer to output layer is
\begin{equation}
\hat{x} = g_{\Theta_{2}} (h) = \sigma(W_{2}h + b_{2})
\end{equation}

Given that the value of each site $x_{i,j}$ of the PCP and the PCPD configurations is "1" (a given site is occupied) or "0" (a given site is not occupied). Thus, the mean cross-entropy over all sites and samples is employed as the loss function for the autoencoder,

\begin{equation}
    L_{loss} = \frac{1}{m}\sum\limits_{r}^{m} \left(\frac{1}{L^2}\sum_{i,j}\left( x_{i,j}\log \hat{x}_{i,j} + (1-x_{i,j})  \log (1-\hat{x}_{i,j})\right) \right),
\end{equation}
where $m$ is the number of samples used in training, and $\hat{x}_{i,j}$, the output of the decoder with $x_{i,j}$ as the input.

Encoders and decoders will be selected as parametric functions (usually neural networks) and will be differentiable to the loss function. So by using a stochastic gradient descent algorithm, the autoencoder can optimize the parameters of the encoder or decoder function.

The parameters of our convolutional autoencoder network are as follows. Three convolutional pooling layers and one fully connected layer are used in the encoder process. In the first convolution layer, $16$ filters are used, the size of the convolution kernel is $2 \times 3$, and the corresponding stride is $1$. The padding form is 'same' to keep the size of the feature map after the convolution operation. Similarly, the second and third convolutions use $8$ filters. All pooling layers use max-pooling, the corresponding filter size is $2 \times 2$, and the stride $2$. In the decoder, the up-sampling layer replaces the pooling layer in the encoder, and it upsamples low-dimensional data to high-dimensional data through a deconvolution filter whose size is $2 \times 2$. The Sigmoid is used as the activation function after each convolutional operation.

\begin{table*}[h]
	\centering
	\begin{tabular}{|c|c|c|}
        \hline
hyperparameters & values  \\
  \hline
learning rate  & 0.0001  \\
        \hline
regularization parameter  & 0.001  \\
       \hline
 batch size & 24  \\
     \hline
  hidden layers of CNN &  2 convolutional and pooling layers, 1 fully connected layer\\
     \hline
 hidden units of FCN &  1 or 2 \\
     \hline
 activation function &  sigmoid \\
     \hline
   \end{tabular}
\caption{The hyper-parameters tuned of autoencoders.}
\label{parameters_ae}
\end{table*}

By minimizing the mean cross-entropy, we obtained detailed information about the hyper-parameters used in autoencoder, please refer to Tab. \ref{parameters_ae}. We chose such an architecture because we have successfully used it in our previous paper \cite{shen2021supervised} to calculate the critical points of another non-equilibrium phase transition model, the directed percolation (DP).

The enormous potential of unsupervised learning has made it popular in scientific research recently. In statistical physics, for systems with tremendous data, unsupervised ML algorithms can process them well. Especially in phase transitions, the Monte Carlo simulations can generate the configuration data of the equilibrium or non-equilibrium phase transition models, and we can use unsupervised ML methods to capture the underlying information of the original data. Therefore, the powerful ability of data processing by unsupervised learning will bring new vitality to the research of statistical physics, which is also the current frontier research hotspot.

\bibliography{sample}

\section*{Acknowledgements}
This work was supported in part by the Fundamental Research Funds for the Central Universities, China (Grant No. CCNU19QN029), the National Natural Science Foundation of China (Grants No.11505071, 61702207, and 61873104), Key Laboratory of Quark and Lepton Physics (MOE), Central China Normal University(Grant No. QLPL2022P01), and the Programme of Introducing Talents of Discipline to Universities (sponsored by the State Administration of Foreign Experts Affairs and the Ministry of Education, PRC) under the 111 Project 2.0 with Grant No. BP0820038.

\section*{Author contributions statement}
Shiyang Chen, Jianmin Shen and Feiyi Liu wrote the Machine learning algorithms, Shengfeng Deng and Dian Xu verified the Monte Carlo simulations, Jianmin Shen and Wei Li computed, obtained all the results of the manuscript, Wei Li supervised the completion of this paper and all the revisions. All authors reviewed the manuscript. 

\section*{Data sets}
All data and ML algorithms related to this study is provided in supplementary file. The detailed algorithms of how to generate raw data and implement machine learning are shown in supplementary material, or refer to the code in the GitHub link {https://github.com/ChuckShen/PCPD-code}.

\section*{Additional information}
The authors have no competing interests as defined by Nature Research, or other interests that might be perceived to influence the results and/or discussion reported in this paper.


\end{document}